\documentclass[12pt,a4paper]{article}

\usepackage[utf8]{inputenc}
\usepackage[english]{babel}
\usepackage[margin=3.0cm]{geometry}
\usepackage{amsmath,amsfonts,amssymb}
\usepackage{fancyhdr}
\usepackage{physics}
\usepackage{graphicx}
\usepackage[framemethod=TikZ]{mdframed}
\usepackage{amsthm}
\usepackage{caption}
\usepackage{subcaption}
\usepackage{array,booktabs}
\usepackage{enumitem}
\usepackage{lipsum}
\usepackage{tensor}
\usepackage{cancel}
\usepackage{mathtools, nccmath}
\usepackage{empheq}
\usepackage{physics}
\usepackage[nottoc]{tocbibind}
\usepackage{lastpage}
\usepackage{hyperref}
\usepackage{cite}
\usepackage{tensor}
\usepackage{slashed}
\usepackage{mathrsfs}
\usepackage[vcentermath]{youngtab}
\usepackage{tocloft}

\theoremstyle{definition}

\newcounter{theo}[section]

\usepackage{hyperref}
\hypersetup{colorlinks=true}
\hypersetup{linkcolor=black}
\hypersetup{citecolor=black}
\hypersetup{urlcolor=black}

\allowdisplaybreaks 

\DeclareFontShape{OT1}{cmr}{mx}{n}%
{<->cmr10}{}
\newcommand{\mytitlefont}{\fontseries{mx}\selectfont}
\DeclareMathAlphabet{\titlemath}{OT1}{cmr}{mx}{n}

\begin{document}

%
\begin{titlepage}
\begin{center}
~\\[1.5cm]
%
{\fontsize{27pt}{0pt} \mytitlefont  When  $\mathbb{Z}_2$ one-form symmetry leads to non-invertible axial symmetries}
%
~\\[1.25cm]
Riccardo Argurio and Romain Vandepopeliere
~\\[0.5cm]
{~{\it Physique Th\'eorique et Math\'ematique and International Solvay Institutes\\
Universit\'e Libre de Bruxelles; C.P. 231, 1050 Brussels, Belgium}}

~\\[1.25cm]
			
\end{center}
\noindent 
We study non-abelian gauge theories with fermions in a representation such that the surviving electric 1-form symmetry is $\mathbb{Z}_2$. This includes $SU(N)$ gauge theories with matter in the (anti)symmetric and $N$ even, and $USp(2N)$ with a Weyl fermion in the adjoint, i.e.~${\cal N}=1$ SYM. We study the mixed 't Hooft anomaly between the discrete axial symmetry and the 1-form symmetry and show that when it is non-trivial, it leads to non-invertible symmetries upon gauging the $\mathbb{Z}_2$. The TQFT dressing the non-invertible symmetry defects is universal to all the cases we study, namely it is always a $U(1)_2$ CS theory coupled to the $\mathbb{Z}_2$ 2-form gauge field. We uncover a pattern where the presence or not of non-invertible defects depends on the rank of the gauge group.

\vfill 
\begin{flushleft}
June 2023
\end{flushleft}
\end{titlepage}
%
		
	
\setcounter{tocdepth}{2}
\renewcommand{\cfttoctitlefont}{\large\bfseries}
\renewcommand{\cftsecaftersnum}{.}
\renewcommand{\cftsubsecaftersnum}{.}
\renewcommand{\cftsubsubsecaftersnum}{.}
\renewcommand{\cftdotsep}{6}
\renewcommand\contentsname{\centerline{Contents}}
	
\tableofcontents

\section{Introduction}

Generalized symmetries \cite{Gaiotto:2014kfa} are one handle on describing the low-energy, strong coupling behavior of gauge theories. The latter behavior depends subtly on the global form of the gauge group \cite{Aharony:2013hda}. In establishing the symmetry content of a theory, one has been traditionally instructed to only keep those that do not suffer ABJ-like anomalies, i.e.~mixed anomalies between global and gauged symmetries. As has been recently uncovered \cite{Choi:2021kmx,Kaidi:2021xfk}, in some cases more symmetries than expected survive, however they become {\em non-invertible}. This means that one can define topological defects that implement symmetry transformations on some objects of the theory, but that also act as projectors on other objects of the theory (typically of different dimensionality), and hence do not form a group structure, but a more general fusion category. For a partial list of references, see \cite{Komargodski:2020mxz, Koide:2021zxj, Apruzzi:2021nmk, Roumpedakis:2022aik,Bhardwaj:2022yxj, Hayashi:2022fkw, Arias-Tamargo:2022nlf, Choi:2022zal,Kaidi:2022uux,  Choi:2022jqy,Cordova:2022ieu, Antinucci:2022eat, Bashmakov:2022jtl,   Damia:2022rxw, Damia:2022bcd,Choi:2022rfe, Bhardwaj:2022lsg , Bartsch:2022mpm,Apruzzi:2022rei, Kaidi:2022cpf,Niro:2022ctq, Mekareeya:2022spm,Antinucci:2022vyk,
Chen:2022cyw,Bashmakov:2022uek, Choi:2022fgx, Yokokura:2022alv, Bhardwaj:2022kot, Bhardwaj:2022maz, Bartsch:2022ytj,Antinucci:2022cdi, Carta:2023bqn}, and also \cite{McGreevy:2022oyu,Cordova:2022ruw } for recent overviews. 

Non-invertible symmetries manifest themselves in different situations and have accordingly different properties: their elements can be taken from a continuous set, from a set related to the rational numbers, or from a finite discrete set. 
In this work, we focus on the latter case, where actually the subset of symmetries which are non-invertible eventually reduces to elements of order two (up to the non-invertibility of course).

Our set-up is the one of four-dimensional non-abelian gauge theories with fer\-mions in a specific real representation. We will restrict to one family (or flavor) of fermions, so that at the classical level the only non-baryonic symmetry is an abelian $U(1)$ axial symmetry. At the quantum level, the axial symmetry is broken to a discrete subgroup because of the ABJ anomaly.\footnote{
In the present set up, this subset of transformations is truly the only survivor. For an abelian gauge group, there are additional non-invertible transformations related to the rational angles in between the discrete values \cite{Choi:2022jqy,Cordova:2022ieu}.
}
However, whether these transformations are invertible or not depends on the global structure of the gauge group, i.e.~on the line content of the theory, and the one-form symmetry under which the lines are charged.

Given a non-abelian gauge algebra, and the fermionic representation, one can directly determine the one-form symmetry of the theory. Let us first consider the simply connected gauge group built from the algebra. Its center determines the one-form symmetry in the pure YM case, which acts on Wilson loops in generic representations. In presence of charged matter, some Wilson loops can end on local operators, and thus render trivial the topological symmetry defects acting non-trivially on them \cite{Heidenreich:2021xpr}. Eventually, the one-form symmetry is reduced to the subgroup of the surviving defects. This subgroup is essentially determined by the representation of the matter fields, in particular what is called the $N$-ality in the $\mathfrak{su}(N)$ case. For instance, for $SU(N)$ gauge group, if matter is in the adjoint representation (i.e.~in ${\cal N}=1$ SYM, since a Weyl fermion is enough when the irreducible representation is real), all of the center is preserved and the one-form symmetry is $\mathbb{Z}_N$. Conversely, in presence of matter in the fundamental (a Dirac fermion now), all of the center becomes trivial and there is no one-form symmetry.

Here, we investigate the intermediate case, in particular when the preserved one-form symmetry is $\mathbb{Z}_2$. Note that the order of the one-form symmetry does not scale with the rank of the gauge group, and this will turn out to be the interesting feature of this situation. 

In presence of a non-trivial one-form symmetry, one can contemplate the possibility of gauging it (or a subgroup of it when it is larger than $\mathbb{Z}_2$), thereby going to a different, non-simply connected global variant of the gauge group. However, there can be a mixed anomaly between the remnant discrete axial symmetry and the one-form symmetry. Hence, gauging the latter is usually not harmless. A subset of the axial transformations remains invertible and forms a group, while the others become non-invertible. 

We will consider gauge algebras whose rank scales with $N$, and fermion representations whose index also scales with $N$ (and hence does the order of the discrete axial symmetry), but which allow for a $\mathbb{Z}_2$ one-form symmetry. Upon gauging the latter, we will see that the presence or not of non-invertible symmetries depends on the details of the theory. In this spirit, we are going to consider non-supersymmetric theories based on the gauge algebra $\mathfrak{su}(N)$ and with matter in the rank-2 tensor representations, as well as ${\cal{N}}=1$ SYM theories based on gauge algebras $\mathfrak{usp}(N)$ and $\mathfrak{so}(N)$. 
In principle, the fate of the discrete axial symmetry upon gauging the $\mathbb{Z}_2$ one-form symmetry is independent of the IR properties of the gauge theory. However, it will turn out that in all the theories that we consider it is reasonable to assume that there is a discrete set of gapped vacua (the number of which scales with the rank), where the remaining discrete axial symmetry (i.e.~the R-symmetry in the supersymmetric case) is spontaneously broken by a fermionic bilinear condensate. As a consequence, we will naturally formulate the action of the symmetries as action on the vacua, and the symmetry defects will be embodied by the domain walls between the vacua.\footnote{Of course, domain walls are physical objects rather than topological defects. Our working assumption is that at low enough energies, only the topological sector of the theory living on the domain wall is relevant to us, and it coincides with the topological theory that defines the defect.}

The results of our investigation are organized as follows. In Section \ref{sec:prelim}, we first quickly review established facts concerning the line content for various global variants of the gauge group: $SU(N)$,  $PSU(N)=SU(N)/\mathbb{Z}_N$ and $SU(N)/\mathbb{Z}_q$ for $q$ a divisor of $N$. We then proceed to discuss the discrete axial symmetry, and its mixed anomaly with the one-form electric symmetry when the gauge group is $SU(N)$. While this is standard when the one-form symmetry is $\mathbb{Z}_N$, i.e.~in ${\cal{N}}=1$ SYM, we notice that the anomaly depends more subtly on the matter representation in the general non-supersymmetric case, i.e.~when it is such that the electric one-form symmetry is originally reduced to $\mathbb{Z}_q$.

For a more in-depth discussion of the vacua and the non-invertible symmetries, we first focus in Section \ref{sec:sun} on $\mathfrak{su}(N)$ gauge algebra and matter in the anti-symmetric or symmetric representation, so that the remaining one-form symmetry is $\mathbb{Z}_2$, assuming $N$ is even. We will first discuss in detail the cases of $\mathfrak{su}(6)$ and $\mathfrak{su}(8)$, underscoring the differences (non-invertible symmetries are present in $SU(6)/\mathbb{Z}_2$ and absent in $SU(8)/\mathbb{Z}_2$). We then briefly consider the case of $\mathfrak{su}(4)$ with a Weyl fermion in the (now real) anti-symmetric representation, before proceeding to considering the general case, which splits into the $\mathfrak{su}(4n+2)$ and $\mathfrak{su}(4n)$ options, with and without non-invertible symmetries respectively, upon gauging the $\mathbb{Z}_2$ one-form electric symmetry.
In all the above cases, when non-invertible symmetry defects are present, they square to an invertible symmetry defect. The TQFT dressing the non-invertible defects turns out to be universal, namely the ${\cal A}^{2,1}$ 3d TQFT \cite{Hsin:2018vcg}. The need for the presence of non-invertible symmetries can be diagnosed by the simultaneous presence, or not, of vacua which are trivially gapped with ones which have a $\mathbb{Z}_2$ gauge theory (i.e., which display confinement or not, respectively).

We then consider in Section \ref{sec:usp} ${\cal N}=1$ SYM with $\mathfrak{usp}(2N)$ gauge algebra. The vacuum structure of these theories is somewhat more familiar, however we point out that when the gauge group is $USp(2N)/\mathbb{Z}_2$ again the presence or absence of non-invertible symmetries depends on $N$ being odd or even, respectively. Also, the non-invertible defects are dressed by the same ${\cal A}^{2,1}$ TQFT. We finally briefly address in Section \ref{sec:spin} theories with $\mathfrak{so}(N)$ gauge algebra, with both odd and even $N$, and then conclude with a small comment on gauging the $\mathbb{Z}_4$ one-form symmetry in $\mathfrak{so}(4n+2)$ SYM.

\section{Preliminaries for \texorpdfstring{$\mathfrak{su}(N)$}{su(N)} gauge algebra}
\label{sec:prelim}

In this work, we consider four-dimensional non-abelian gauge theories with a single massless fermion in a self-conjugate representation $R$ of the gauge group. More specifically, we take a left Weyl fermion $\psi$ in an irreducible representation $\rho$ along with another left Weyl fermion $\tilde{\psi}$ transforming in $\rho^*$, so that in total the self-conjugate reducible representation is $R = \rho \oplus \rho^*$. One sees them together as a single Dirac fermion $\Psi$. If the irreducible representation is real $\rho = \rho^*$, a single Weyl fermion $\lambda$ is enough. With such precautions, 
we avoid gauge anomalies.

This section concerns gauge theories that share the same Lie algebra $\mathfrak{su}(N)$, and hence that are locally equivalent. We review how the line operator content of these theories can differ \cite{Aharony:2013hda}, as well as the mixed 't Hooft anomaly between the axial (discrete) zero-form symmetry for the fermion and the electric one-form symmetry for $SU(N)$ gauge group. This mixed axial-electric 't Hooft anomaly is the starting point for the study of the non-invertible symmetry defects.

\subsection{Line operator content of global variants}  

\subsubsection{ \texorpdfstring{$SU(N)$}{SU(N)}}
Pure Yang-Mills $SU(N)$ gauge theory
\begin{equation}
S = \int \Tr( -\frac{1}{2g^2} f \wedge \star f + \frac{\theta}{8\pi^2} f \wedge f ) \, 
\end{equation}
has a $\mathbb{Z}_N^{(1)}$ electric one-form symmetry (which is the center symmetry) and no magnetic one-form symmetry. The only line operators allowed in the theory are purely electric lines, with charges $(n,0)$ for $n= 0,1, ... , N-1$, and where $n$ is the $N$-ality of the representation attached to the line. The $\mathbb{Z}_N^{(1)}$ electric one-form symmetry acts on these Wilson lines as 
\begin{equation}
(n,0) \rightarrow e^{\frac{2\pi i kn}{N}} (n,0) \, , 
\end{equation}
for $k=0,1,...,N-1$. The shift $ \theta \rightarrow \theta + 2 \pi$ leaves the theory manifestly invariant since the instanton number is quantized.\footnote{In this paper we always assume that the spacetime manifold has a spin structure.} It also maps the line operators to themselves, since they do not carry any magnetic charge. Hence, $ \theta \rightarrow \theta + 2 \pi$ is a genuine symmetry of the theory and $\theta$ is $2\pi$-periodic.

If one adds a fermion in an arbitrary representation $R$ of the gauge group $SU(N)$, only a subgroup of the center will survive as electric symmetry, namely the subgroup that does not act on the charged matter. Let us denote $c$ the $N$-ality of $R$. Among the $N$ elements of the center subgroup $\mathbb{Z}_N = \{ z_n \, I_N = e^{2\pi i n /N} I_N \, | \, \, \, n=0,1,...,N-1 \} \subset SU(N)$ where $I_N$ is the $N \times N$ unit matrix, only $q = \textrm{gcd}(N,c)$ of them are mapped to the identity by the homomorphism $R$, since $R [z_n] = z_n^{c}$. They are given by 
\begin{equation}\label{eq:ZqDefinitionSubgroup}
\mathbb{Z}_q = \{ e^{\frac{2\pi i k}{N}}  I_N \, , \, \, \, k = 0, a, 2a,... \, , (q-1)a    \}
\end{equation}
where $a$ is defined such that $N=aq$.

\subsubsection{\texorpdfstring{$SU(N)/\mathbb{Z}_N = PSU(N)$}{PSU(N)}}

There is no electric one-form symmetry in the $PSU(N)$ gauge theory, but it has a $\mathbb{Z}_N^{(1)}$ magnetic one-form symmetry acting on magnetically charged lines \cite{Aharony:2013hda}.

All the purely electric lines (i.e. with $m = 0$) should have $n$ multiple of $N$ and hence they are in the class $(0, 0)$, so that there is no non-trivial purely electric line. If one takes a line with minimal magnetic charge, in the class of $(n,1)$, then lines $(n',1)$ with $ n' \neq n$ are not allowed due to the Dirac quantization condition
\begin{equation}\label{eq:DiracQuant}
nm' - mn' = 0 \, \,  \textrm{mod} \, N
\end{equation}
but lines of the class $(n',m)$ are allowed if $n' = nm \, \, \textrm{mod} \, N$. All in all, we end up with the set of lines 
\begin{equation}\label{eq:PSUNlines}
L_n = \{  (nm,m) \, \, \textrm{mod} \, N \, \, |\, \,  m = 0,1,..., N-1 \} \,
\end{equation}
which means that we have a distinct line operator content for every $n=0,1,...,N-1$, hence $n$ distinct theories sharing the same $PSU(N)$ gauge group. We shall denote the theories by 
\begin{equation}
PSU(N)_n = \left( \frac{SU(N)}{\mathbb{Z}_N} \right)_n  \, \, \, \textrm{with} \, \, \, n = 0,1,..., N-1 \, ,
\end{equation}
whose line operators have charges in $L_n$. For instance, the theory with $n=0$ exhibits purely magnetic lines only, while the other ones contain dyonic lines. 
Let us emphasize that these $N$ inequivalent theories differ globally but the local physics remains the same in all of them.

Now of course, the transformation $\theta \rightarrow \theta + 2\pi$ is not a genuine symmetry anymore due to the Witten effect. Indeed, the shift $(nm,m) \rightarrow ( (n+1)m , m)$ maps the sets $L_n$ to $L_{n+1}$, hence takes us from the $PSU(N)_n$ to the $PSU(N)_{n+1}$ theory. The general rule is that the theories are related to each other by the $\theta$-angle shift as 
\begin{equation}
PSU(N)_{n}^{\theta + 2 \pi k} = PSU(N)_{(n+k) \, \, \textrm{mod} \, N}^\theta \,  
\end{equation}
so that the periodicity of $\theta$ obviously becomes $\theta \sim \theta + 2 \pi N$, in order to end up on the same theory.

\subsubsection{\texorpdfstring{$SU(N)/\mathbb{Z}_q$}{SU(N)/Zq}}\label{sec:SUN/Zq}

Here we describe the case in which the electric one-form symmetry in the $SU(N)$ theory is broken to $\mathbb{Z}_q^{(1)}$ by the presence of fermions.
By definition, $q$ is a divisor of $N$, so that we can define the integer $a$ such that $N=aq$ as in (\ref{eq:ZqDefinitionSubgroup}). If one gauges the electric $\mathbb{Z}_q^{(1)}$ in $SU(N)$, the only one-form symmetry that remains in the $SU(N)/\mathbb{Z}_q$ gauge theory is a magnetic $\mathbb{Z}_q^{(1)}$ symmetry.\footnote{For a complete analysis of  $SU(N)/\mathbb{Z}_q$ in the pure gauge case, see \cite{Gaiotto:2014kfa,Aharony:2013hda}. We will not need it in the following.}

The purely electric line operators are of the form $(mq,0)$ mod $N$, and hence are all trivial since they can end on local (fermionic) operators. The line operator with minimal magnetic charge is $(n,a)$ mod $N$ for some $n = 0,1,..., q-1$. 
The set of non-trivial lines is then
\begin{equation}
L_n^{q} = \{  m (n,a) \, \, \textrm{mod} \, \, N \, \, | \, \, m \in \mathbb{Z}    \, \, \textrm{and} \, \, n = 0 ,1 , ..., q-1       \} \, . 
\end{equation}
Clearly, it matches with (\ref{eq:PSUNlines}) if one takes $q=N$, hence $a = 1$. At the end, there are $q$ distinct theories 
\begin{equation}
\left(   \frac{SU(N)}{\mathbb{Z}_q} \right)_n \, \, \, \textrm{with} \, \, n=0,1,..., q-1
\end{equation}
that all share the same gauge group but that have $L_n^{q}$ as line content.

A $2 \pi$ shift of $\theta$ is generally not a genuine symmetry of the theory. It transforms $(n,a) \rightarrow (n+a,a)$, and maps $L_n^{q}$ to $L_{n+a}^{q}$, hence one has 
\begin{equation}\label{eq:WittenFroSUZq}
\left( \frac{SU(N)}{\mathbb{Z}_q} \right)_n^{\theta + 2\pi} = \left( \frac{SU(N)}{\mathbb{Z}_q} \right)_{(n+a) \, \, \textrm{mod} \, q}^{\theta}\ .
\end{equation}
Something peculiar might happen here. If gcd$(q,a) = 1$, i.e.~if they are coprime, every choice of $n$ can be reached by shifting the $\theta$-angle. However, if it is not the case, some theories cannot be connected via a shift of $\theta$. More precisely, defining gcd$(q,a) = p \neq 1$, the shift of $\theta$ only maps the theory $(SU(N) / \mathbb{Z}_q)_n$ to other theories with the same $n$ mod $p$. This implies there are $p$ sets of theories that are not related by shifts of the $\theta$-angle.  This can happen if and only if $N$ has some prime factor that appears more than once in its decomposition into primes.

\subsection{Mixed axial-electric 't Hooft anomaly}\label{sec:AxialEleAnomaly}

We note $T_a$ with $a = 1,..., N^2-1$ the generators of the Lie algebra $su(N)$, normalized such that
\begin{equation}
\Tr [T_a T_b] = \frac{1}{2} \delta_{ab}\ ,
\end{equation}
where the trace is in the fundamental representation. The Dynkin index $l(R)$ of a representation $R$ is defined here by 
\begin{equation}\label{eq:Dynkinindex}
\Tr_R [T_a T_b] = l(R) \Tr [T_a T_b] \, . 
\end{equation}
If a representation $R$ is the direct sum of two representations $R = \rho_1 \oplus \rho_2$, then its Dynkin index is $l(R)= l(\rho_1)+ l(\rho_2)$. In $SU(N)$, the adjoint representation has $l = 2N$, the rank-2 antisymmetric representation has $l=N-2$ and the rank-2 symmetric one has $l=N+2$.

The axial symmetry  $U(1)^{(0)}_A : \Psi \rightarrow e^{i \alpha \gamma_5} \Psi $ is quantum mechanically broken by an ABJ anomaly \cite{PhysRev.177.2426,Bell1969APP}. By Fujikawa's analysis \cite{PhysRevLett.42.1195}, one learns that under a chiral rotation of angle $\alpha$, the path-integral measure changes as 
\begin{equation}
\mathcal{D}\Psi \, \mathcal{D} \bar{\Psi} \rightarrow \mathcal{D}\Psi \,  \mathcal{D} \bar{\Psi} \exp \left(     i  \frac{  \alpha }{8 \pi^2} \int \Tr_R (f \wedge f             )     \right) = \mathcal{D}\Psi \,  \mathcal{D} \bar{\Psi} \exp \left(     i  \frac{  \alpha l }{8 \pi^2} \int \Tr (f \wedge f             )     \right) \, , 
\end{equation}
which means that only a subgroup $\mathbb{Z}_l^{(0)}$ survives in the quantum theory, namely the rotations of angle
\begin{equation}
\alpha = \frac{2\pi n}{l} \,  , \, \, \,\textrm{with} \, \, \, n \in \{0,1,...,l-1\} \, . 
\end{equation}
Here $l$ is the total Dynkin index of the reducible representation, i.e.~it would be $l=2$ for a Dirac fermion in the fundamental, and $l=2(N\pm2)$ for a Dirac fermion in the (anti)symmetric.

Note that there is an equivalence between a non-anomalous axial rotation of angle $\alpha = 2\pi n/l  $ and the transformation in the space of couplings $\theta \rightarrow \theta + 2\pi n$, directly at the level of the path integral.

The story however does not end here.
There is a mixed 't Hooft anomaly between the axial zero-form symmetry and the electric one-form symmetry \cite{Gaiotto:2014kfa}. This mixed anomaly is related to the fact that the instanton number can become fractional in presence of the two-form electric background field $B_e$. We discuss this first for the case of the adjoint representation and then for an arbitrary representation.


We begin with $R = Adj$, realizing $\mathcal{N}=1$ supersymmetry, in order to have $\mathbb{Z}_N^{(1)}$ as electric one-form symmetry. It is more convenient to embed our $SU(N)$ gauge bundle into $U(N)$ in order to couple the theory to the background field. To do so, we add a non-vanishing trace to the $SU(N)$ gauge connection $a$ in the following way 
\begin{equation}
\tilde{a} = a_a T_a + a' \mathbb{I}_N \, ,
\end{equation}
where $\tilde{a}$ and $a'$ are respectively $U(N)$ and $U(1) \subset U(N)$ gauge fields. 

One introduces the electric background field $B_e$ which has holonomies in $\mathbb{Z}_N $, i.e. $\frac{1}{2\pi} \int B_e \in \frac{1}{N} \mathbb{Z}$, as
\begin{equation}
\Tr \tilde{f} = N B_e 
\end{equation}
so that one expresses $\tilde{f}$ as the following composition of the $SU(N)$ field strength and the background field
\begin{equation}
\tilde{f} = f + \frac{1}{N} \Tr \tilde{f} \,  \mathbb{I}_N = f + B_e  \, \mathbb{I}_N \, . 
\end{equation}
Now, the instanton number becomes
\begin{align}
\nu =  \frac{1}{8\pi^2}\int \Tr (f \wedge f             )  = \frac{1}{8\pi^2} \int \Tr ( \tilde{f} \wedge \tilde{f} ) - \frac{N}{8\pi^2} \int  B_e \wedge B_e 
\end{align}
and, by using the fact that the second Chern class $ c_2 = \frac{1}{8\pi^2} ( \Tr \tilde{f} \wedge \tilde{f} - \Tr \tilde{f} \wedge \Tr \tilde{f} )$ of $U(N)$ is integral, $\int c_2 \in \mathbb{Z}$, one writes the fractional contribution of $B_e$ to the instanton number as
\begin{align}
\nu &= \mathbb{Z}  + \frac{1}{8\pi^2} \int \Tr \tilde{f} \wedge \Tr \tilde{f} - \frac{N}{8\pi^2} \int B_e \wedge B_e   
= \mathbb{Z} +\frac{N(N-1)}{8\pi^2} \int B_e \wedge B_e  \, . 
\end{align}
Since $B_e$ describes a $\mathbb{Z}_N$ bundle, one can use a 2-cocycle representative $\hat{B}_e \in H^2(X,\mathbb{Z}_N)$ that satisfies $\oint \hat{B}_e \, \in \{0,1,..., N-1\} = \mathbb{Z} \mod N$, defined via $ N B_e = 2\pi \hat{B}_e$. At the end, we are left with 
\begin{equation}
\nu = 
\mathbb{Z} + \frac{N-1}{2N} \int \mathcal{P}[\hat{B}_e] \, ,  
\end{equation}
in terms of the Pontryagin square\footnote{The Pontryagin square is a cohomology operation $\mathcal{P} : H^2(X,\mathbb{Z}_N) \rightarrow H^4(X,\mathbb{Z}_{\textrm{gcd}(N,2) \, N})$.
It is an even integer class on spin manifolds, which will always be the case for us since we have fermions in the theory. Hence we can always take it as $\mathcal{P}[\hat{B}_e]=  \hat{B}_e \cup \hat{B}_e$.} of the background field.
Eventually, recalling that the Dynkin label of the adjoint representation is $ l = 2N$, the mixed 't Hooft anomaly between an axial rotation of angle $\alpha = 2\pi n / 2N$ and the electric one-form symmetry for a fermion in the adjoint is given by 
\begin{equation}\label{eq:MixedAnomSUn}
\frac{Z[B_e; \, \theta + 2\pi n]}{Z[B_e; \,  \theta]} = e^{iA} \, \, \, \, \, \,\, \, \,\, \, \,\, \, \, \textrm{with} \, \, \, \, \, \,\, \, \,\, \, \,\, \, \, A =  \frac{2\pi n(N-1)}{2N} \int  \mathcal{P}[\hat{B}_e] \, . 
\end{equation}
The integral of the Pontryagin square is an even number, so that among the axial rotations $n \in \{0,1,...,2N-1\}$, only $n \in \{0, N\}$ remain non-anomalous in presence of the electric background field. Hence, only a $\mathbb{Z}_2^{(0)}$ subgroup\footnote{This is actually fermion parity.} of $\mathbb{Z}_{2N}^{(0)}$ does not suffer from the mixed axial-electric 't Hooft anomaly and will survive if one gauges the electric one-form symmetry.


If the fermion is in an arbitrary representation $R$ of $N$-ality $c$, the electric background field will now have holonomies in $\mathbb{Z}_q$, i.e. $\frac{1}{2\pi} \int B_e \in \frac{1}{q} \mathbb{Z}$, with $q=$gcd$(N,c)$. When coupling this background to the theory via the $U(N)$ bundle, one must now impose 
\begin{equation}
\Tr \tilde{f} = q B_e \, , 
\end{equation}
so that $\Tr \tilde{f}$ still has proper $U(1)$ holonomies $\frac{1}{2\pi} \int \Tr \tilde{f} \in \mathbb{Z}$. This implies the decomposition
\begin{equation}
\tilde{f} = f + \frac{1}{N} \Tr \tilde{f} \,  \mathbb{I}_N = f + \frac{q}{N} B_e \, \mathbb{I}_N \, , 
\end{equation}
and, following the same lines, the instanton number becomes 
\begin{align}
\nu &= \frac{1}{8\pi^2}\int \Tr ((\tilde{f} - \frac{q}{N} B_e \mathbb{I}_N) \wedge (\tilde{f} - \frac{q}{N} B_e \mathbb{I}_N)            )
= \mathbb{Z} + \frac{q^2}{8\pi^2} \left( \frac{N-1}{N} \right) \int  B_e \wedge B_e \, . 
\end{align}
Going into the 2-cocycle description $qB_e = 2\pi \hat{B}_e$, one finds
\begin{equation}
\nu =  
\mathbb{Z} + \frac{N-1}{2N} \int \mathcal{P}[\hat{B}_e] \, ,  
\end{equation}
which is the same form as the result for the adjoint representation.

However, note that while the prefactors are the same as for the adjoint, the cocycles have different natures, hence the anomalies are not equivalent. Indeed, as defined in (\ref{eq:ZqDefinitionSubgroup}), the 2-cocycle $\hat{B}_e$ is such that its integral over a 2-cycle takes value in the $\mathbb{Z}_q$ subgroup of $\mathbb{Z}_N$, namely
\begin{equation}
\int \hat{B}_e \in \{ 0, a , 2a, ... , (q-1) a  \} \, , 
\end{equation}
where $a = N/q$. To deal with a genuine $\mathbb{Z}_q$-valued cocycle, we define $\hat{B}_e'$ such that $\hat{B}_e = a \hat{B}_e'$ and the anomalous phase becomes
\begin{align}\label{eq:MixedAnomGenSUn}
A = \frac{2\pi n(N-1)}{2N} \int  \mathcal{P}[\hat{B}_e] 
&= \frac{2\pi n(N-1)a}{2q}  \int  \mathcal{P}[\hat{B}_e'] 
= \frac{2\pi n(N-1)a'}{2q'}  \int  \mathcal{P}[\hat{B}_e'] \, , 
\end{align}
where we have taken into account that the fraction $a/q$ could be reducible. Then one defines $a/q = a'/q'$ where $a'$ and $q'$ are coprime, i.e.~gcd$(a',q')=1$. The integral of the Pontryagin squared is an even number, so that the mixed 't Hooft anomaly is eventually valued in $\mathbb{Z}_{q'}$. 
Among all the values that $n=0,1,.., l-1$ can take, the only rotations that will leave the partition function invariant will by characterized by the $n$'s that are integer multiple of $q'$. Hence only $\mathbb{Z}_{l/q'} \subset \mathbb{Z}_l$ remains non-anomalous, in the sense that it does not have a mixed 't Hooft anomaly with the electric one-form symmetry.\footnote{In all cases that one can check, it is easy to convince oneself that $q'$ divides $l$. It is not obvious to show this in all generality by algebraic methods, however our arguments provide a physics proof of sorts.
} Note that if one gauges this electric symmetry, i.e.~if one makes the electric background gauge field dynamical by summing over it in the path-integral, one will end up with a $\mathbb{Z}_{l/q'}^{(0)}$ axial symmetry for the fermion in the gauged theory.

\section{Non-invertible symmetries in \texorpdfstring{$SU(N)/\mathbb{Z}_2$}{SU(N)/Z2} with 2-index anti-symmetric representation}
\label{sec:sun}

We focus now on four-dimensional $SU(N)$ gauge theory with one flavour of fermions in the anti-symmetric $\tiny\yng(1,1)$ representation.\footnote{A very similar theory has been recently considered in \cite{Anber:2023pny}, focusing though on different properties.} It has $N$-ality equal to 2, i.e. $c=2$ in our notation, so that we will restrict $N$ to be even in order for $q=$ gcd$(N,c)$ to be different than 1, and by extension to have an electric one-form symmetry $\mathbb{Z}_2^{(1)}$ in our theory. Excluding $SU(2)$, we consider $N \geq 4$.

Note that we focus on the anti-symmetric representation $\tiny\yng(1,1)$, but there is a similar story for the symmetric representation $\tiny\yng(2)$, as we will comment briefly in  Sec.~\ref{sec:symmetric}.
The only exception is  for $SU(4)$, where the anti-symmetric representation is real and we will take one Weyl fermion only.

Let us quickly remind the form of the beta function in a generic theory.
For a gauge group $G$ and $n_f$ Dirac fermions in a representation $R_f$ of $G$, the one-loop beta function for the coupling constant $g$ is
\begin{equation}
\beta (g) = - \left( \frac{11}{3} C_2 (G) - \frac{1}{3} n_f \, l(R_f) \right) \frac{g^3}{16 \pi^2} \, , 
\end{equation}
where $C_2(G)$ is the quadratic Casimir of $G$ and $l(R_f)$ is the Dynkin index. The quadratic Casimir of the gauge group is equal to one half of the Dynkin index of the adjoint representation. The quadratic Casimir of $SU(N)$ is given by $N$, whereas $l(\tiny\yng(1,1)) = 2(N-2)$ and $n_f = 1$. It means that 
\begin{equation}
\beta (g) = - \left(3 N+ \frac{4}{3} \right)\frac{g^3}{16 \pi^2} \, , 
\end{equation}
and the theory will be asymptotically free for any $N$.\footnote{This property ceases to be true for matter in representations of $N$-ality higher than 2. Indeed their Dynkin index scales in $N$ with a power higher than linear.} We assume that at low-energies, the theory undergoes confinement into gapped vacua, where moreover a bilinear fermionic condensate forms, breaking the discrete axial symmetry $\mathbb{Z}^{(0)}_{2(N-2)}$ to $\mathbb{Z}^{(0)}_2$.\footnote{In principle, this assumption could be relaxed since the axial-electric mixed anomaly is at most $\mathbb{Z}_2$-valued, hence no more than 2 confining vacua would actually be enough. This would mean that $\langle \psi \tilde{\psi} \rangle =0$ while a higher-order condensate possibly forms, namely $\langle ( \psi \tilde{\psi})^{(N-2)/2} \rangle \neq 0$. However, the fact that a bilinear fermionic condensate nevertheless forms can be supported by arguments such as large $N$ equivalence with ${\cal N}=1$ SYM \cite{Armoni:2003gp}, and some refined 't Hooft anomaly matching involving also the baryonic symmetry \cite{Anber:2021lzb,Tanizaki:2022plm}. See \cite{Yamaguchi:2018xse} for a slightly more contrived example of a gauge theory where only a higher-order condensate forms.}

The goal of this section is to gauge the $\mathbb{Z}_2^{(1)}$ and to study the resulting $SU(N)/ \mathbb{Z}_2$ gauge theory with one fermion in the anti-symmetric representation. We start with $SU(6)$ and $SU(8)$, and generalize the results to $SU(4n +2)$ and $SU(4n)$. In the $SU(N)/ \mathbb{Z}_2$ theories, non-invertible axial symmetry defects separate two adjacent vacua that are physically inequivalent for $N = 4n+2$, while all the vacua remain physically equivalent and there are no non-invertible symmetry defects for $N = 4n$.
We also treat the special case of $SU(4)$ on its own, which has one vacuum only and  does not contain any non-invertible symmetry defect.

\subsection{\texorpdfstring{$SU(6) / \mathbb{Z}_2$}{SU(6)}}

\subsubsection{Before gauging}

The theory $SU(6) + \tiny\yng(1,1)$ has a $\mathbb{Z}_8^{(0)}$ axial symmetry and a $\mathbb{Z}_2^{(1)}$ electric symmetry. Since $a = 3$ and $q=2$, they are coprime and $q'=q$.

In the infrared, one assumes the fermion bilinear $\tilde{\psi} \psi$ has a non-vanishing vacuum expectation value, i.e.~it condenses. This spontaneously breaks $\mathbb{Z}_8^{(0)} \rightarrow \mathbb{Z}_2^{(0)}$, resulting in 4 isolated vacua that are related by the broken generators of the coset $\mathbb{Z}_4^{(0)} = \mathbb{Z}_8^{(0)} / \mathbb{Z}_2^{(0)}$. More precisely, they correspond to the first four rotations of the group. If one writes generically the group as
\begin{equation}
\mathbb{Z}_8 = (1, z , z^2, z^3,z^4,z^5,z^6,z^7) \, ,
\end{equation}
the group that rotates the vacua into each other is $\mathbb{Z}_4^{(0)} = (1, z , z^2, z^3)$. Indeed, the $\mathbb{Z}_2^{(0)}$ that remains in the infrared is composed of $1$ and $z^4$.
From another point of view,
the rotation to move from the $j^{\textrm{th}}$ to the $(j+k)^{\textrm{th}}$ vacuum is equivalent to the shift $\theta \rightarrow \theta + 2\pi k$. Since the latter is a genuine symmetry of the theory, all the 4 vacua are physically equivalent. 
Looking at (\ref{eq:MixedAnomGenSUn}) with $q'=2$, we deduce that among the 8 axial rotations only $(1,z^2,z^4,z^6)$ do not suffer from the mixed axial-electric 't~Hooft anomaly.

What about the line operators? The pure $SU(6)$ Yang-Mills theory has purely electric lines only and they are simply given by
\begin{equation}
SU(6) : \{ (0,0), (1,0),(2,0),(3,0),(4,0),(5,0)  \} \, . 
\end{equation}
Taking the antisymmetric fermionic matter into account, the $\mathbb{Z}_6^{(1)}$ electric symmetry is now broken to $\mathbb{Z}_2^{(1)}$, so that 
\begin{equation}
SU(6) +  \textrm{$\tiny\yng(1,1)$} \,: \{ (0,0), (1,0) \} \, . 
\end{equation}

\subsubsection{After gauging : $SU(6)/ \mathbb{Z}_2 $ }

Now we gauge $\mathbb{Z}_2^{(1)}$ and the theory on which we end up is $SU(6)/ \mathbb{Z}_2 +  \tiny\yng(1,1)$ . It has a $\mathbb{Z}_2^{(1)}$ magnetic one-form symmetry and $\mathbb{Z}_4^{(0)}$ as an axial symmetry.

What is the line content for the pure gauge theory? 
Actually, there are two distinct gauge theories that share the $SU(6)/\mathbb{Z}_2$ gauge group, as reviewed in Sec.~\ref{sec:SUN/Zq}. They have the same local physics but differ in their line operators content. In other words, they are not globally equivalent. Concretely, $(SU(6)/\mathbb{Z}_2)_0$ contains integer linear combinations of $(2,0)$ and $(0,3)$, while $(SU(6)/\mathbb{Z}_2)_1$ is made of integer linear combinations of $(2,0)$ and $(1,3)$. Hence, one has
\begin{equation}
\left( \frac{SU(6)}{\mathbb{Z}_2} \right)_0 : \{ (0,0),(2,0),(4,0),(0,3),(2,3),(4,3) \} \, , 
\end{equation}
\begin{equation}
\left( \frac{SU(6)}{\mathbb{Z}_2} \right)_1 : \{ (0,0),(2,0),(4,0),(1,3),(3,3),(5,3) \} \, .
\end{equation}
From (\ref{eq:WittenFroSUZq}), we know that the two theories must be related by a $\theta$-shift since gcd$(a,q) = 1$. It is indeed the case due to the Witten effect: performing $\theta \rightarrow \theta + 2\pi$ on the lines of the first theory generates the lines of the second theory, and vice versa. Note that both these pure gauge theories have a $\mathbb{Z}_6^{(1)}=\mathbb{Z}_2^{(1)}\times \mathbb{Z}_3^{(1)}$ one-form symmetry.

In order to take the fermion into account in these theories, one has to mod out the electric charges by 2. It gives 
\begin{equation}
\left( \frac{SU(6)}{\mathbb{Z}_2} \right)_0 +  \textrm{$\tiny\yng(1,1)$} \, : \{ (0,0),(0,3) \} \, , 
\end{equation}
\begin{equation}
\left( \frac{SU(6)}{\mathbb{Z}_2} \right)_1 +  \textrm{$\tiny\yng(1,1)$} \,: \{ (0,0), (1,3) \} \, .
\end{equation}
Both theories have a $\mathbb{Z}_{2,m}^{(1)}$ one-form symmetry that plays no role in the following.

\subsubsection{Vacuum structure}\label{sec:SU6VacStruc}

\begin{figure}
	\center
	\captionsetup{justification=centering,margin=2cm}
	\includegraphics[scale=0.7]{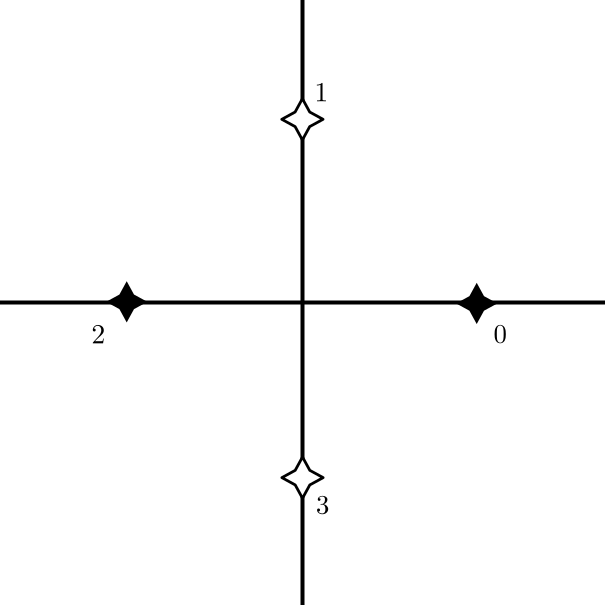}
	\caption{Vacuum structure of $SU(6)$ + AntiSym \\
	Filled points : $(0,1)$ condenses in this vacuum.\\
	Unfilled points : $(1,1)$ condenses in this vacuum.\\
	All vacua are trivially gapped.}
	\label{fig:4vacuaSU6}
\end{figure}

In the $SU(6) +  \textrm{$\tiny\yng(1,1)$}$ theory, one can assume confinement in the IR is due to the condensation of a dyonic particle in each vacuum. We expect that all vacua will be trivially gapped, so that the low-energy physics is the same in each of them, as expected from the fact that the $\theta$-shift is a symmetry in the theory. A good candidate to start with is the pure 't Hooft line $(0,1)$. Likewise, the condensing line in the next vacuum reached via $\theta \rightarrow \theta + 2 \pi$ is $(1,1)$. It turns back to $(0,1)$ after a new $2\pi$-shift of $\theta$, since the electric charges are defined modulo 2. Let us assume that $(0,1)$ condenses in the $0^\textrm{th}$ vacuum and the rest will follow via the Witten effect. One has eventually that the dyon $(k \, \textrm{mod } 2, 1)$ condenses in the $k^\textrm{th}$ vacuum, or in other words :
\begin{align}
\textrm{Vacua 0,2} & \rightarrow \langle T_{(0,1)} \rangle \neq 0  \, , \nonumber\\
\textrm{Vacua 1,3} & \rightarrow \langle T_{(1,1)} \rangle \neq 0 \, .
\end{align}
The vacuum structure can be seen in Fig.~\ref{fig:4vacuaSU6}. None of the genuine lines that are available in this theory is aligned with the condensing dyons. Indeed, the only non-trivial line is purely electric and has $(1,0)$ charge. Consequently, in all the vacua the electric one-form symmetry remains unbroken and all 4 vacua are trivially gapped, as expected. However, the axial-electric mixed anomaly allows us to further characterize the vacua by a non-trivial $\mathbb{Z}_2$ valued symmetry protected topological phase (SPT), i.e.~a $\hat{B}_e'$-dependent phase:
\begin{align}
    \mathrm{SPT}_{k}= e^{\frac{\pi i}{2}k\int \mathcal{P}(\hat{B}_e')}\ .
\end{align}
We can assign vacua 0 and 2 the SPT$_0$, while the other two will have SPT$_1$.

The $SU(6)/\mathbb{Z}_2 +  \textrm{$\tiny\yng(1,1)$}$ theory has 4 distinct vacua, as the ungauged theory. However, two of the four transformations that rotate the vacua into each other have now become non-invertible. Indeed, among the four rotations $(1, z , z^2, z^3)$, one knows that $(z,z^3)$ suffer from the mixed 't Hooft anomaly with the electric one-form symmetry, so that gauging the latter implies them to become non-invertible. Concretely, in this theory  
\begin{equation}
\theta \rightarrow \theta + 2\pi k \, \,   \left\{
    \begin{array}{ll}
        \textrm{is invertible} & \textrm{for }  k=0,2 \\
        \textrm{is non-invertible} & \textrm{for }  k=1,3 \, . 
    \end{array}
\right.
\end{equation}

Since $\theta \rightarrow \theta + 4 \pi$ is a genuine invertible symmetry of the theory, the vacua are 2 by 2 physically equivalent. We expect to find the same low energy physics in the vacua 0 and 2, and the same for vacua 1 and 3. 
This is a new feature compared to $\mathcal{N}=1$ SYM $PSU(N)$.\footnote{
In $\mathcal{N}=1$ $PSU(N)$ SYM, the vacua are indeed inequivalent: for instance for $N$ prime one finds that $N-1$ of them are trivially gapped but with a distinct SPT, while one hosts a $\mathbb{Z}_N$ gauge theory \cite{Choi:2022zal}. For general $N$ there are also vacua with both an SPT and a $\mathbb{Z}_n$ gauge theory, where $n$ divides $N$, see e.g.~\cite{Damia:2023ses}.} Let us confirm this explicitly by checking the fate of the line operators within each vacuum.

\begin{figure}
     \centering
     \begin{subfigure}[b]{0.45\textwidth}
         \centering
         \includegraphics[width=\textwidth]{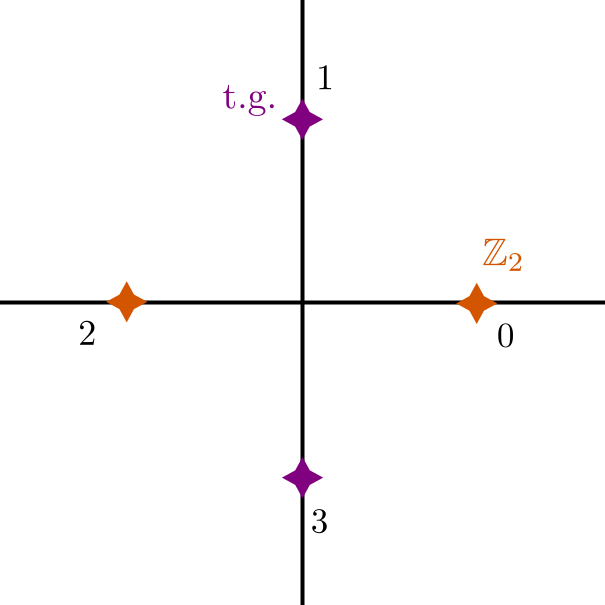}
         \caption{$ (SU(6)/ \mathbb{Z}_2)_0 +  \textrm{$\tiny\yng(1,1)$}$}
         \label{fig:4vacuaSU6Z20}
     \end{subfigure}
     \hfill
     \begin{subfigure}[b]{0.45\textwidth}
         \centering
         \includegraphics[width=\textwidth]{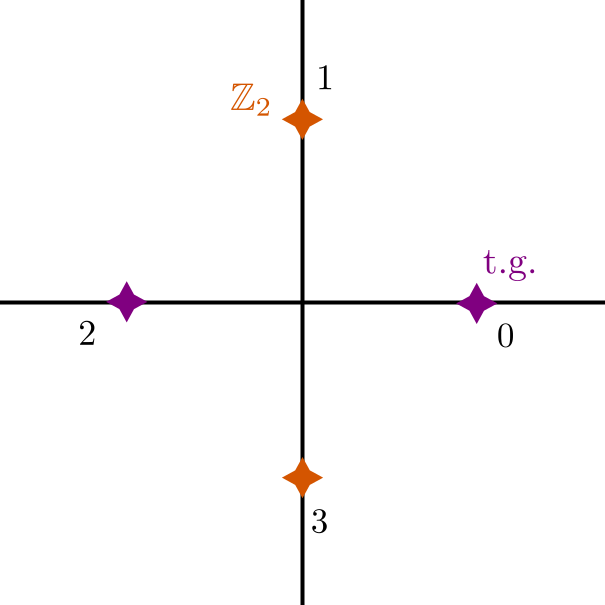}
         \caption{$ (SU(6)/ \mathbb{Z}_2)_1 +  \textrm{$\tiny\yng(1,1)$}$}
         \label{fig:4vacuaSU6Z21}
     \end{subfigure}
     	\captionsetup{justification=centering,margin=2cm}
        \caption{Vacuum structure of the two gauged theories. \\
        Purple : Trivially gapped vacua. Orange : $\mathbb{Z}_2$ gauge theory.}
        \label{fig:4vacuaSU6Z2}
\end{figure}

In each vacuum, the dyonic particle that condenses must be the same as in the $SU(6)$ theory. Now, the genuine line operators that are available in the theory are no longer purely electric and can be aligned to the condensing dyon. Since two line operators that are in the same equivalence class have the same IR behaviour, a genuine line of the theory can now follow a perimeter law and thus break the one-form symmetry. Let us treat the two inequivalent theories in turn: 
\begin{itemize}
\item $\left( \frac{SU(6)}{\mathbb{Z}_2} \right)_0 +  \textrm{$\tiny\yng(1,1)$}$ only has $(0,3)$ that carries non-trivial charges. In vacua 0 and 2, it is aligned with the condensing line $(0,1)$ because $(0,3) \sim (0,1)^3$, hence it also exhibits a perimeter law signaling the complete breaking of the magnetic $\mathbb{Z}_2^{(1)}$ one-form symmetry in these vacua. This results in an emergent $\mathbb{Z}_2$ magnetic gauge symmetry. All in all, the IR theory in these two vacua is the same, namely a $\mathbb{Z}_2$ gauge theory. However, it still has an area law in the vacua 1 and 3, where the low energy theory is trivially gapped.
\begin{align}
\textrm{Vacua 0,2} & \rightarrow \, \, \,  \langle T_{(0,3)} \rangle \neq 0  \, \,  \, \rightarrow \, \mathbb{Z}_2 \, \,  \textrm{gauge theory}  \, , \nonumber\\
\textrm{Vacua 1,3} & \rightarrow \textrm{area law only}  \rightarrow \textrm{trivially gapped} \, .
\end{align}
\item $\left( \frac{SU(6)}{\mathbb{Z}_2} \right)_1 +  \textrm{$\tiny\yng(1,1)$}$ only has $(1,3)$ that carries non-trivial charges. In vacua 0 and 2, it still has an area law, and the low energy theory is trivially gapped. It is aligned with the condensing $(1,1)$ line in vacua 1 and 3 because $(1,3) \sim (1,1)^3$, resulting in a $\mathbb{Z}_2$ gauge theory.
\begin{align}
\textrm{Vacua 0,2} & \rightarrow \textrm{area law only}  \rightarrow \textrm{trivially gapped} \, , \nonumber\\
\textrm{Vacua 1,3} & \rightarrow \, \, \,  \langle T_{(1,3)} \rangle \neq 0  \, \,  \, \rightarrow \, \mathbb{Z}_2 \, \,  \textrm{gauge theory}  \, . 
\end{align}
\end{itemize}
One can see the result of this analysis in Fig.~\ref{fig:4vacuaSU6Z2}.
The consequence of a $2\pi$-shift of $\theta$ is two-fold : on one hand, it goes from one vacuum to the next one and on the other hand, it goes from one theory to the other theory by exchanging the genuine line operators. Hence, the vacua in $(SU(6)/\mathbb{Z}_2)_0$ are physically equivalent to those in $(SU(6)/\mathbb{Z}_2)_1$, provided that one relabels them. Indeed, vacua 0 \& 2 of $(SU(6)/\mathbb{Z}_2)_0$ are equivalent to vacua 1 \& 3 in $(SU(6)/\mathbb{Z}_2)_1$ and the same goes for two remaining pairs. 

Note that the trivially gapped vacua actually all carry an SPT$_1$. Indeed, the above vacuum structure can be determined from the gauging of the vacuum SPTs of $SU(6)$. For $(SU(6)/\mathbb{Z}_2)_0$, gauging SPT$_0$ gives the $\mathbb{Z}_2$ gauge theory, while gauging SPT$_1$ gives back the same SPT$_1$. For $(SU(6)/\mathbb{Z}_2)_1$, one has to stack an SPT$_1$ before gauging, so that the same results hold up to a rotation among the vacua. For a more in-depth discussion of this approach, see \cite{Damia:2023ses}.

\subsubsection{Domain walls}\label{sec:DomainWallsSU6}

The domain wall between each adjacent pair of vacua in $ SU(6)/ \mathbb{Z}_2 +  \textrm{$\tiny\yng(1,1)$}$ should contain a 3d TQFT in order to compensate for the different physics on its two sides. Another way to see this is to start with the symmetry defect in the ungauged theory. In the more general $SU(N) + \Psi$ theory, if one takes the domain wall between two successive vacua, the two sides will differ by the anomalous term 
\begin{equation}
\frac{2\pi (N-1)}{2} \frac{a'}{q'} \int \mathcal{P}[\hat{B}_e'] \, ,
\end{equation}
obtained from (\ref{eq:MixedAnomGenSUn}). Now, note that if $q'=q$ both $\hat{B}_e'$ and the mixed anomaly are valued in $\mathbb{Z}_{q}$. 
Upon gauging the one-form symmetry, one should stack the 3d minimal abelian TQFT $\mathcal{A}^{q,a(N-1)}$ on the domain wall in order to compensate the anomaly. Indeed, recall \cite{Hsin:2018vcg} that an $\mathcal{A}^{N,p}$ theory has a $\mathbb{Z}_N^{(1)}$ one-form symmetry whose anomaly is
\begin{equation}\label{eq:Rappel}
-2\pi \frac{p}{2N} \int \mathcal{P}[\hat{B}] \, ,
\end{equation}
where $\hat{B}$ is the $\mathbb{Z}_N$ 2-cocycle. 
For our current theory, one has $q=2$ and $a(N-1) = 15$, but since $p$ is defined mod $N$ on spin manifolds, the 3d TQFT that needs to decorate this basic domain wall is 
\begin{equation}
\mathcal{A}^{2,1} = U(1)_2 \, . 
\end{equation}
Upon gauging the 1-form symmetry (i.e.~making $\hat{B}_e'$ dynamical), this makes the domain wall no longer invertible. Similarly, the conjugate domain wall is decorated with $\overline{\mathcal{A}}^{2,1} = \mathcal{A}^{2,-1} \simeq  \mathcal{A}^{2,1}$, since $U(1)_2$ is time-reversal invariant as a spin-TQFT \cite{Hsin:2016blu}. As explained in \cite{Kaidi:2021xfk}, the product theory $\mathcal{A}^{2,1} \times \mathcal{A}^{2,-1}$ is equivalent to a $\mathbb{Z}_2$ Dijkgraaf-Witten theory with a non-trivial twist, that we will call the condensation defect $\cal C$. The basic domain wall $\mathcal{N}$ separating two adjacent vacua consists of the $\mathbb{Z}_8$ axial symmetry defect $D$ inserted on $M_3$ decorated with the topological theory $\mathcal{A}^{2,1}$, in the sense that
\begin{align}
\mathcal{N} (M_3) &= D(M_3) \mathcal{A}^{2,1}(M_3) \nonumber\\
&\sim D(M_3) \, \int \mathcal{D}a \,   \exp( i \frac{2}{4\pi} \int_{M_3} a d a - \frac{i}{2\pi} \int_{M_3} a {b}_e') 
\end{align}
where $a$ is the $U(1)$ dynamical gauge field that leaves on the 3d defect only and is coupled through the last term to the $\mathbb{Z}_2$ discrete 2-form gauge field $b_e'$, which is gauging the electric 1-form symmetry. The axial symmetry defect/operator is nothing but the ordinary exponential of $\star j_A$ integrated over $M_3$, with angles taking value in $\mathbb{Z}_8$. It is invertible in the usual sense $D \times \bar{D} =1$ and is such that $D \times D = D^2$, $ D^2 \times D^2 =1$.\footnote{$D$ is effectively a generator of $\mathbb{Z}_4$ because the $\mathbb{Z}_2$ subgroup of $\mathbb{Z}_8$ is actually part of the gauge symmetry, since $-1\in SU(6)$.} Of course, $\bar{D} = D^3$. One has 
\begin{align}
\mathcal{N} (M_3) \times \overline{\mathcal{N}} (M_3) &= \mathcal{C} (M_3)\ .
\end{align}
It is clearly not invertible anymore, since $\cal C$ can be seen as a sum over Wilson surfaces of $b_e'$ \cite{Kaidi:2021xfk}. One also has 
\begin{equation}
\mathcal{N} (M_3) \times \mathcal{N} (M_3) = D^2(M_3) \, \mathcal{C} (M_3)\, . 
\end{equation}
Note that the domain wall to go from vacuum 0 to vacuum 2 remains invertible since it does not have to host any TQFT. Hence, it simply consists of $D^2(M_3)$. Let us denote it $\mathcal{N}_2 (M_3)$, then one has
\begin{equation}
\mathcal{N} (M_3) \times \mathcal{N} (M_3) =  \mathcal{C}(M_3) \, \mathcal{N}_2 (M_3)\ ,
\end{equation}
i.e.~two non-invertible defects fuse non-trivially into an invertible defect.

\subsubsection{What changes for matter in the symmetric}\label{sec:symmetric}
We here briefly state what happens when the fermion is in the symmetric representation rather the antisymmetric. $SU(6)+\tiny\yng(2)$ has a $\mathbb{Z}_{16}^{(0)}$ axial symmetry, which is again broken to $\mathbb{Z}_{2}^{(0)}$ by the fermionic bilinear condensate. There are hence 8 vacua. The line content and the anomaly are exactly the same as before. Therefore, in the $SU(6)/\mathbb{Z}_2+\tiny\yng(2)$ theories, the vacua again alternate between those which are trivially gapped and those that host a $\mathbb{Z}_2$ gauge theory. The domain walls are the same as above, the only difference being in the invertible part which now is such that $D^8=1$, taking into account the gauge symmetry.

\subsection{\texorpdfstring{$SU(8)/ \mathbb{Z}_2$}{SU(8)}}
\subsubsection{Before gauging}
The theory $SU(8) + \tiny\yng(1,1)$ has a $\mathbb{Z}_{12}^{(0)}$ axial symmetry and a $\mathbb{Z}_2^{(1)}$ electric symmetry. Since $a = 4$ and $q=2$, they are not coprime and $a/q = 2 = a'/q'$ with $q'=1$.

In the infrared, the fermion bilinear condenses $\langle \psi \tilde{\psi} \rangle \neq 0$, yielding the spontaneous symmetry breaking $\mathbb{Z}_{12}^{(0)} \rightarrow \mathbb{Z}_{2}^{(0)}$. There are 6 isolated vacua that are related by the broken generators of $\mathbb{Z}_{6}^{(0)} = \mathbb{Z}_{12}^{(0)} / \mathbb{Z}_{2}^{(0)}$, or equivalently by shifts of $\theta$. These are symmetries of the theory and the vacua are all physically equivalent. Since $q'=1$, there is no axial-electric mixed 't Hooft anomaly.

What about the line operators? The pure $SU(8)$ Yang-Mills theory only has purely electric lines that are  
\begin{equation}
SU(8) : \{ (0,0), (1,0),(2,0),(3,0), (4,0),(5,0),(6,0),(7,0) \} \, . 
\end{equation}
Taking the antisymmetric fermionic matter into account, the $\mathbb{Z}_8^{(1)}$ electric symmetry is now reduced to $\mathbb{Z}_2^{(1)}$,  so that there are only two inequivalent lines 
\begin{equation}
SU(8) +  \textrm{$\tiny\yng(1,1)$} \,: \{ (0,0), (1,0)  \} \, . 
\end{equation}

\subsubsection{After gauging}

Now we gauge $\mathbb{Z}_2^{(1)}$ and the theory on which we end up is $SU(8)/ \mathbb{Z}_2 +  \tiny\yng(1,1)$ . It has a $\mathbb{Z}_2^{(1)}$ magnetic one-form symmetry and still $\mathbb{Z}_{12}^{(0)}$ as an axial symmetry.

What is the line content for the pure gauge theory? 
Again, there are two distinct gauge theories that share the $SU(8)/\mathbb{Z}_2$ gauge group, as we know from Sec.~\ref{sec:SUN/Zq}. Concretely, $(SU(8)/\mathbb{Z}_2)_0$ contains integer linear combinations of $(2,0)$ and $(0,4)$, while $(SU(8)/\mathbb{Z}_2)_1$ is made of integer linear combinations of $(2,0)$ and $(1,4)$. Hence, one has
\begin{equation}
\left( \frac{SU(8)}{\mathbb{Z}_2} \right)_0 : \{ (0,0),(2,0),(4,0),(6,0),(0,4), (2,4), (4,4),(6,4) \} \, , 
\end{equation}
\begin{equation}
\left( \frac{SU(8)}{\mathbb{Z}_2} \right)_1 : \{ (0,0),(2,0),(4,0),(6,0),(1,4), (3,4), (5,4),(7,4)\} \, .
\end{equation}
From (\ref{eq:WittenFroSUZq}), we know that the two theories cannot be related by a $\theta$-shift since gcd$(a,q) \neq 1$. More specifically, one sees that a $2\pi$-shift of $\theta$ leaves the line content of each theory invariant. Within each theory, the $\theta$-shift is a genuine symmetry, differently from the $SU(6)$ case. 
Note that these pure gauge theories differ by their 1-form symmetry: $(SU(8)/\mathbb{Z}_2)_0$ has $\mathbb{Z}_{4}^{(1)} \times \mathbb{Z}_{2}^{(1)}$ while $(SU(8)/\mathbb{Z}_2)_1$ has $\mathbb{Z}_8^{(1)}$ (see  \cite{Gaiotto:2014kfa} and also \cite{Bergman:2022otk}).

Now we take the fermion into account, this yields the electric charges to be modded out by 2. It gives 
\begin{equation}
\left( \frac{SU(8)}{\mathbb{Z}_2} \right)_0 +  \textrm{$\tiny\yng(1,1)$} \, : \{ (0,0),(0,4) \} \, , 
\end{equation}
\begin{equation}
\left( \frac{SU(8)}{\mathbb{Z}_2} \right)_1 +  \textrm{$\tiny\yng(1,1)$} \,: \{ (0,0), (1,4) \} \, .
\end{equation}

\subsubsection{Vacuum structure}

\begin{figure}
	\center
	\captionsetup{justification=centering,margin=2cm}
	\includegraphics[scale=0.7]{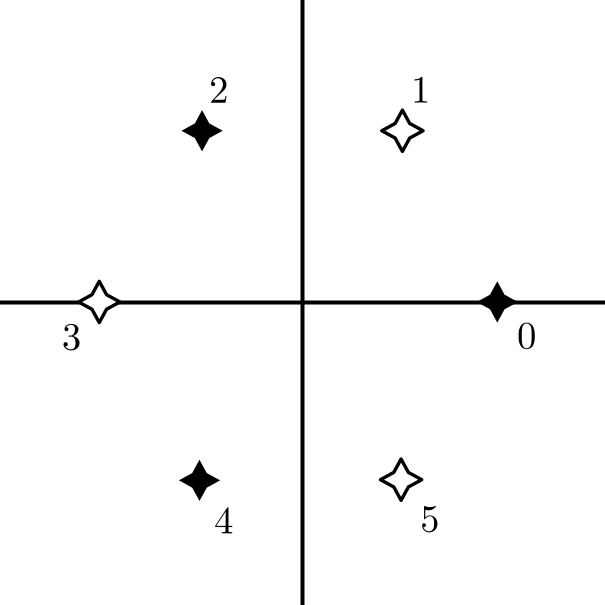}
	\caption{Vacuum structure of $SU(8)$ + antiSym \\
	Filled dots : $(0,1)$ condenses in this vacuum\\
	Unfilled dots : $(1,1)$ condenses in this vacuum}
	\label{fig:6vacuaSU8}
\end{figure}

The $SU(8) +  \textrm{$\tiny\yng(1,1)$}$ theory is assumed to be confining in the IR.
Let us assume the confinement at low energies is due to the condensation of a dyonic particle in each vacua. As before, a good candidate for this is the $(0,1)$ line, which is a pure 't Hooft magnetic line. It is not a genuine line of the theory. Under $\theta \rightarrow \theta + 2\pi$, the condensing line turns into $(1,1)$ and then the latter goes back to $(0,1)$ after a new $2\pi$-shift of $\theta$. Since the vacua are related by the symmetries $\theta \rightarrow \theta + 2 \pi k$, we assume that $(0,1)$ condenses in the $0^{\textrm{th}}$ vacuum and the rest will follow via the Witten effect:
\begin{align}
\textrm{Vacua 0,2,4    } & \rightarrow \langle T_{(0,1)} \rangle \neq 0  \, , \nonumber\\
\textrm{Vacua 1,3,5    } & \rightarrow \langle T_{(1,1)} \rangle \neq 0 \, .
\end{align}
This vacuum structure is depicted in Fig.\ref{fig:6vacuaSU8}. 
All 6 vacua are trivially gapped and are totally equivalent, even at the level of the partition function since they all have a trivial SPT phase, due to the absence of a mixed 't Hooft anomaly.

\begin{figure}
     \centering
     \begin{subfigure}[b]{0.45\textwidth}
         \centering
         \includegraphics[width=\textwidth]{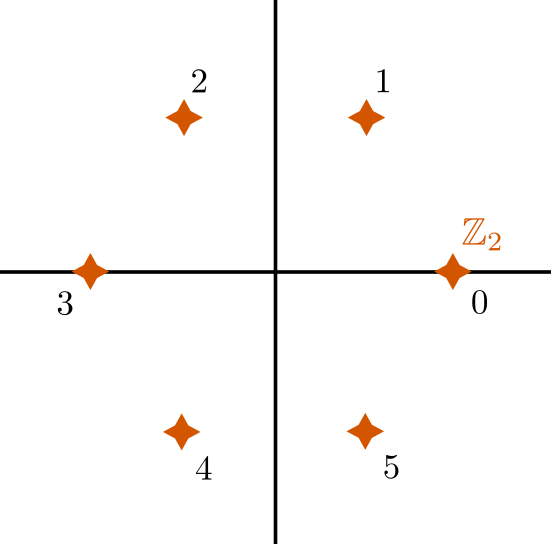}
         \caption{$ (SU(8)/ \mathbb{Z}_2)_0 +  \textrm{$\tiny\yng(1,1)$}$}
         \label{fig:6vacuaSU8Z20}
     \end{subfigure}
     \hfill
     \begin{subfigure}[b]{0.45\textwidth}
         \centering
         \includegraphics[width=\textwidth]{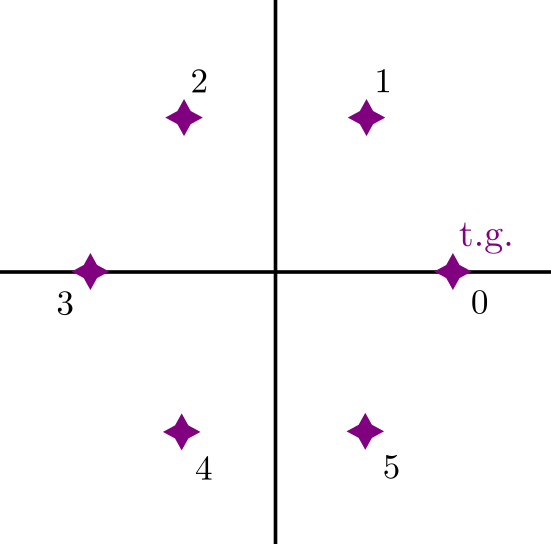}
         \caption{$ (SU(8)/ \mathbb{Z}_2)_1 +  \textrm{$\tiny\yng(1,1)$}$}
         \label{fig:6vacuaSU8Z21}
     \end{subfigure}
     	\captionsetup{justification=centering,margin=2cm}
        \caption{Vacuum structure of the two gauged theories. \\
        Purple : Trivially gapped vacua. Orange : $\mathbb{Z}_2$ gauge theory.}
        \label{fig:6vacuaSU8Z2}
\end{figure}

In the $SU(8)/\mathbb{Z}_2 +  \textrm{$\tiny\yng(1,1)$}$ theories, there must still be 6 vacua and the same dyonic particle as in the $SU(8)$ case that condenses in each one of them. Since there was no mixed 't Hooft anomaly in the ungauged theory, the transformations that go from one vacuum to another, i.e. $2\pi$-shifts of $\theta$, are still genuine invertible symmetries and a $\mathbb{Z}_6^{(0)}$ still connects the 6 vacua to each other.
Now the genuine line operators of the theory are no longer purely electric, hence can be aligned with the dyon that condenses in each vacuum. 
However, we know that the physics must be the same in all 6 vacua of the theory, since they are related to each other by a symmetry. One has to treat the two inequivalent theories separately: 
\begin{itemize}
\item $\left( \frac{SU(8)}{\mathbb{Z}_2} \right)_0 +  \textrm{$\tiny\yng(1,1)$}$ only has $(0,4)$ that carries non-trivial charges. This line is aligned with both condensing lines $(0,1)$ and $(1,1)$ in the sense that $(0,4) = (0,1)^4$ and $(0,4) = (1,1)^4$, since the electric charge is mod-2 defined. Hence, in each of the 6 vacua, the IR theory is a $\mathbb{Z}_2$ gauge theory. 
\item $\left( \frac{SU(8)}{\mathbb{Z}_2} \right)_1 +  \textrm{$\tiny\yng(1,1)$}$ only has $(1,4)$ that carries non-trivial charges, but it is never aligned with any condensing candidate in any vacuum. The magnetic one-form symmetry remains unbroken and all 6 vacua are trivially gapped.
\end{itemize}
In both theories, the 6 vacua are physically equivalent.\footnote{The fact that the vacua of $( {SU(8)}/{\mathbb{Z}_2} )_0 +  \textrm{$\tiny\yng(1,1)$}$ all carry a $\mathbb{Z}_2$ gauge theory can also be seen by gauging the trivial SPT in the vacua of $SU(8) +  \textrm{$\tiny\yng(1,1)$}$. As for $( {SU(8)}/{\mathbb{Z}_2} )_1 +  \textrm{$\tiny\yng(1,1)$}$, one has to stack an SPT$_1$ before gauging, so that all vacua eventually carry the same SPT$_1$ also after gauging.} This also confirms the fact that a $\theta$-shift is a symmetry for each theory, that goes from one vacuum to another, while staying within a given theory. It can be seen in Fig.\ref{fig:6vacuaSU8Z2}. There is no non-invertible symmetry in any of the two theories.

The same conclusions can be drawn when the fermion is in the symmetric representation, with the only difference being that the number of vacua is 10.

\subsection{\texorpdfstring{$SU(4)/ \mathbb{Z}_2$}{SU(4)}}

Recall that this theory is special in the sense that the antisymmetric representation is real, hence a single Weyl fermion can carry it. This is what we will suppose in this subsection. Note then that there is no analog with the symmetric representation.

\subsubsection{Before gauging}
The theory $SU(4) + \tiny\yng(1,1)$ has a $\mathbb{Z}_2^{(0)}$ axial symmetry and a $\mathbb{Z}_2^{(1)}$ electric symmetry. Since $a = 2$ and $q=2$, they are not coprime and $a/q = 1 = a'/q'$ with $q'=1$.

In the infrared, there is no constraint on the fermion bilinear vacuum expectation value. The axial symmetry remains unbroken and there is only one vacuum. 
Since $q'=1$, there is no mixed 't Hooft anomaly between the axial zero-form symmetry and the electric one-form symmetry. This agrees with the assumption of a single vacuum.

What about the line operators? The pure $SU(4)$ Yang-Mills theory only has purely electric lines that are simply 
\begin{equation}
SU(4) : \{ (0,0), (1,0),(2,0),(3,0)  \} \, . 
\end{equation}
Taking the antisymmetric fermionic matter into account, the $\mathbb{Z}_4^{(1)}$ electric symmetry is now reduced to $\mathbb{Z}_2^{(1)}$,  so that there are only two inequivalent lines 
\begin{equation}
SU(4) +  \textrm{$\tiny\yng(1,1)$} \,: \{ (0,0), (1,0)  \} \, . 
\end{equation}

\subsubsection{After gauging}

Now we gauge $\mathbb{Z}_2^{(1)}$ and the theory on which we end up is $SU(4)/ \mathbb{Z}_2 +  \tiny\yng(1,1)$ . It has a $\mathbb{Z}_2^{(1)}$ magnetic one-form symmetry and still $\mathbb{Z}_2^{(0)}$ as an axial symmetry.

What is the line content for the pure gauge theory? 
Again, there are two distinct gauge theories that share the $SU(4)/\mathbb{Z}_2$ gauge group. Concretely, $(SU(4)/\mathbb{Z}_2)_0$ contains integer linear combinations of $(2,0)$ and $(0,2)$, while $(SU(4)/\mathbb{Z}_2)_1$ is made of integer linear combinations of $(2,0)$ and $(1,2)$. Hence, one has
\begin{equation}
\left( \frac{SU(4)}{\mathbb{Z}_2} \right)_0 : \{ (0,0),(2,0),(0,2),(2,2) \} \, , 
\end{equation}
\begin{equation}
\left( \frac{SU(4)}{\mathbb{Z}_2} \right)_1 : \{ (0,0),(2,0),(1,2),(3,2) \} \, .
\end{equation}
The two theories are not related by a  $2\pi$-shift of $\theta$, which is then a genuine symmetry.

With the fermionic matter, one is left with 
\begin{equation}
\left( \frac{SU(4)}{\mathbb{Z}_2} \right)_0 +  \textrm{$\tiny\yng(1,1)$} \, : \{ (0,0),(0,2) \} \, , 
\end{equation}
\begin{equation}
\left( \frac{SU(4)}{\mathbb{Z}_2} \right)_1 +  \textrm{$\tiny\yng(1,1)$} \, : \{ (0,0),(1,2) \} \, .
\end{equation}

\subsubsection{Vacuum structure}

Again, one assumes $SU(4) +\textrm{$\tiny\yng(1,1)$}$ confines in the IR and that it is due to the condensation of a dyonic/magnetic particle in the vacuum. Note that here the line must be mapped to itself under a $2\pi$-shift of $\theta$ so that one remains in the one and only vacuum. We must then exclude that the monopole of charge $(0,1)$ condenses. The dyon of charge $(1,1)$ is rejected for the very same reason.
Thus one is led to assume that it is $(0,2)$ that condenses in the vacuum 
\begin{equation}
\langle T_{(0,2)} \rangle \neq 0 \, . 
\end{equation}
Of course the purely electric line $(1,0)$ is not influenced by this condensation, so that we end up with a trivially gapped theory at low energies in the vacuum.

In the two $SU(4)/\mathbb{Z}_2 +\textrm{$\tiny\yng(1,1)$}$ theories, one still has one vacuum only, characterized by the condensation of the 't Hooft line $(0,2)$ as in the ungauged theory. Since the two theories are not related by a $\theta$-shift, one expect their respective vacua to be physically different: 

\begin{itemize}
\item $\left( \frac{SU(4)}{\mathbb{Z}_2} \right)_0 +  \textrm{$\tiny\yng(1,1)$}$ only has $(0,2)$ that carries non-trivial charges. This line is nothing but the condensing line, so that the IR theory is a $\mathbb{Z}_2$ gauge theory in the vacuum.
\item $\left( \frac{SU(4)}{\mathbb{Z}_2} \right)_1 +  \textrm{$\tiny\yng(1,1)$}$ only has $(1,2)$ that carries non-trivial charges, but it is not aligned with the condensing line. Hence, the magnetic one-form symmetry remains unbroken and the vacuum is trivially gapped.\footnote{For the same reasons as in the $SU(8)$ case, we expect an SPT$_1$ in this vacuum.}
\end{itemize}

Finally, let us remark that $SU(4)/\mathbb{Z}_2+ \textrm{$\tiny\yng(1,1)$}$ is nothing else than $  SO(6) + \textrm{$\tiny\yng(1)$}$. The latter theory also comes in two versions differing in the line content.

\subsection{\texorpdfstring{$SU(4n+2)/ \mathbb{Z}_2$}{SU(4n+2)/Z2} for \texorpdfstring{$n \geq 1$}{n>=1}}
Let us generalize the previous analysis for all $\mathfrak{su}(N)$ algebras, with even $N$. We start with $\mathfrak{su}(4n+2)$ with $n \geq 1$, which as we will see reflects perfectly the $\mathfrak{su}(6)$ case, up to the total number of vacua.

\subsubsection{Before gauging}
The theory $SU(4n+2) + \textrm{$\tiny\yng(1,1)$}$ has a $\mathbb{Z}_{8n}^{(0)}$ axial symmetry and a $\mathbb{Z}_2^{(1)}$ electric symmetry. Since $q =2$ and $N=aq$, one has $a= 2n+1$ so that $a$ and $q$ are relatively prime and $q'=q = 2$.

At low energies, one assumes that the fermion bilinear condenses $\langle \tilde{\psi} \psi \rangle \neq 0$ and this yields the spontaneous breaking of the axial symmetry $\mathbb{Z}_{8n}^{(0)} \rightarrow \mathbb{Z}_{2}^{(0)}$. The IR theory has $4n$ isolated vacua, that are related by the broken generators of $\mathbb{Z}_{4n}^{(0)} = \mathbb{Z}_{8n}^{(0)}/ \mathbb{Z}_{2}^{(0)}$. Actually, these $4n$ transformations are the first $4n$ rotations of $\mathbb{Z}_{8n}^{(0)}$. From another point of view, the rotation to move form the $j^{\textrm{th}}$ to the $(j+k)^{\textrm{th}}$ vacuum is equivalent to the shift $\theta \rightarrow \theta + 2\pi k$. Since the latter is a genuine symmetry of the theory, all the $4n$ vacua are physically equivalent. 
Looking at (\ref{eq:MixedAnomGenSUn}) with $q'=2$, we deduce that among the $8n$ axial rotations only the even rotations $(1,z^2,z^4,...,z^{8n-2})$ do not suffer from the mixed axial-electric 't Hooft anomaly. All in all, the characteristics of the axial rotations are 
\begin{equation}\label{eq:Z8nAxial}
\mathbb{Z}_{8n} = ( \underbrace{\mathbf{1}, z ,\mathbf{z^2},z^3,\mathbf{z^4}, ..., z^{4n-1} }_{\textrm{rotate the } 4n \textrm{ vacua}} , \mathbf{z^{4n}},z^{4n+1},..., \mathbf{z^{8n-2}},z^{8n-1} )
\end{equation} 
where the boldface stands for rotations that do not suffer from the mixed 't Hooft axial-electric anomaly.

The line operators of $SU(4n+2) + \textrm{$\tiny\yng(1,1)$}$ are purely electric. Due to the charged matter, the electric charges of the lines are defined mod 2. Thus the set of lines is 
\begin{equation}
SU(4n+2) +  \textrm{$\tiny\yng(1,1)$} \,: \{ (0,0), (1,0) \} \, . 
\end{equation}

\subsubsection{After gauging}

Now we gauge $\mathbb{Z}_2^{(1)}$ and we end up on the theory $SU(4n+2)/\mathbb{Z}_2 +  \textrm{$\tiny\yng(1,1)$}$. It has a $\mathbb{Z}_2^{(1)}$ magnetic one-form symmetry and a $\mathbb{Z}_{4n}^{(0)}$ axial symmetry, which corresponds to the non-anomalous rotations of (\ref{eq:Z8nAxial}).

There are two distinct theories that share this gauge group. For pure gauge theories, $(SU(4n+2)/\mathbb{Z}_2)_0$ contains integer combinations of $(2,0)$ and $(0,2n+1)$, while $(SU(4n+2)/\mathbb{Z}_2)_1$ is made of integer linear combinations of $(2,0)$ and $(1,2n+1)$. Now with the fermions, one simply has : 
\begin{equation}
\left( \frac{SU(4n+2)}{\mathbb{Z}_2} \right)_0 +  \textrm{$\tiny\yng(1,1)$} \, : \{ (0,0),(0,2n+1) \} \, , 
\end{equation}
\begin{equation}
\left( \frac{SU(4n+2)}{\mathbb{Z}_2} \right)_1 +  \textrm{$\tiny\yng(1,1)$} \,: \{ (0,0), (1,2n+1) \} \, .
\end{equation}
As expected from the fact that gcd$(a,q)=$gcd$(2n+1,2)=1$, the two theories are related by $\theta \rightarrow \theta + 2\pi$, in the sense that this shift exchanges the lines of the two theories due to the Witten effect.

\subsubsection{Vacuum structure}

In the $SU(4n+2) +  \textrm{$\tiny\yng(1,1)$}$ theory, we assume that the confinement  at low energies is due to the condensation of a dyonic particle in each vacuum. It must be such that one ends up with every vacuum being trivially gapped. A good candidate to start with is the pure 't Hooft line of charge $(0,1)$. The condensing line in the next vacuum (reached via $\theta \rightarrow \theta + 2 \pi$) is $(1,1)$, and it goes back to $(0,1)$ in the following vacuum. We assume that $(0,1)$ condenses in the $0^{\textrm{th}}$ vacuum and the rest follows through the Witten effect. Eventually, the dyon that condenses in the $k^{\textrm{th}}$ vacuum has charges $(k$ mod $2,1)$. It means 
\begin{align}
\textrm{Vacua } 0, 2,  4, ... ,  4n-2 & \rightarrow \langle T_{(0,1)} \rangle \neq 0 \, , \nonumber\\
\textrm{Vacua } 1,3,  5, ... , 4n-1 & \rightarrow \langle T_{(1,1)} \rangle \neq 0 \, ,
\end{align}
The line operator which is genuine in the theory, namely $(1,0)$ is never aligned with the lines that condense in the vacua. Hence, in all the vacua it exhibits an area law and the electric one-form symmetry remains unbroken. The $4n$ vacua end up being trivially gapped, as expected.\footnote{Similar to the discussion in (\ref{sec:SU6VacStruc}), the vacua actually alternate between SPT$_0$ and SPT$_1$.} They are related by a genuine invertible symmetry and are indeed physically equivalent.

The $SU(4n+2)/\mathbb{Z}_2 +  \textrm{$\tiny\yng(1,1)$}$ theory also has $4n$ distinct vacua. However, one half of the transformations that rotate the vacua into each other have now become non-invertible, namely those that are not in boldface in (\ref{eq:Z8nAxial}) : $(z,z^3,..., z^{4n-1})$. This means that in the theory
\begin{equation}
\theta \rightarrow \theta + 2\pi k \, \,   \left\{
    \begin{array}{ll}
        \textrm{is a symmetry} & \textrm{for  }  k=0,2, ... ,4n-2 \\
        \textrm{is a non-invertible symmetry} & \textrm{for  }  k=1,3, ... , 4n-1 \, . 
    \end{array}
\right.
\end{equation}
In this way, the low energy physics must be equivalent in all vacua that are labeled by an even number, and so it must be the case in all the odd vacua, too. However, these two low energy physics are inequivalent. One can check this explicitly by looking at the fate of the line operators within each vacuum.

In each vacuum, the dyonic particle that condenses must be the same in the $SU(4n+2)$ theory and in the $SU(4n+2)/\mathbb{Z}_2$ theories. Now, the genuine line operators that are available in the theory are no longer purely electric and can be aligned to the condensing dyon. Since two line operators that are in the same equivalence class have the same IR behaviour, a genuine line of the theory can now follow a perimeter law and thus break the one-form symmetry. Let us treat the two inequivalent theories separately: 
\begin{itemize}
\item $\left( \frac{SU(4n+2)}{\mathbb{Z}_2} \right)_0 +  \textrm{$\tiny\yng(1,1)$}$ only has $(0,2n+1)$ that carries non-trivial charges. In the even vacua, it is aligned with the condensing line $(0,1)$ because $(0,2n+1) \sim (0,1)^{2n+1}$, hence it also exhibits a perimeter law signaling the breaking of the magnetic $\mathbb{Z}_2^{(1)}$ one-form symmetry in these vacua. This results in an emergent $\mathbb{Z}_2$ magnetic gauge symmetry. All in all, the IR theory in these $2n$ vacua is the same, namely a $\mathbb{Z}_2$ gauge theory. However, the line $(0,2n+1)$ still has an area law in the odd vacua, where the low energy theory is trivially gapped.
\begin{align}
\textrm{Vacua } 0,2, ... , 4n-2 & \rightarrow  \,  \langle T_{(0,2n+1)} \rangle \neq 0  \,  \rightarrow \, \mathbb{Z}_2 \, \,  \textrm{gauge theory}  \, , \nonumber\\
\textrm{Vacua } 1,3, ... , 4n-1 & \rightarrow \textrm{area law only}  \rightarrow \textrm{trivially gapped} \, .
\end{align}
\item $\left( \frac{SU(4n+2)}{\mathbb{Z}_2} \right)_1 +  \textrm{$\tiny\yng(1,1)$}$ only has $(1,2n+1)$ that carries non-trivial charges. In even vacua, it still has an area law, and the low energy theory is trivially gapped. It is aligned with the condensing $(1,1)$ line in odd vacua because $(1,2n+1) \sim (1,1)^{2n+1}$, resulting in a $\mathbb{Z}_2$ gauge theory.
\begin{align}
\textrm{Vacua } 0,2, ... , 4n-2 & \rightarrow \textrm{area law only}  \rightarrow \textrm{trivially gapped} \, , \nonumber\\
\textrm{Vacua } 1,3, ... ,  4n-1 & \rightarrow \,   \langle T_{(1,2n+1)} \rangle \neq 0  \,  \rightarrow \, \mathbb{Z}_2 \, \,  \textrm{gauge theory}  \, . 
\end{align}
\end{itemize}
The consequence of a $2\pi$-shift of $\theta$ is two-fold : on one hand, it goes from one vacuum to the next one and on the other hand, it goes from one theory to the other theory by exchanging the genuine line operators. Hence, the vacua in $(SU(4n+2)/\mathbb{Z}_2)_0$ are physically equivalent to those in $(SU(4n+2)/\mathbb{Z}_2)_1$, provided that one exchanges the even vacua for the odd vacua. Indeed, the even vacua of $(SU(4n+2)/\mathbb{Z}_2)_0$ are equivalent to the odd vacua in $(SU(4n+2)/\mathbb{Z}_2)_1$ and the same goes for the two complementary sets.\footnote{This is the case even at the level of the partition function, since now one alternates between the $\mathbb{Z}_2$ gauge theory and the SPT$_1$.}

\subsubsection{Domain walls}
The domain wall between each adjacent pairs of vacua in $ SU(4n+2)/ \mathbb{Z}_2 +  \textrm{$\tiny\yng(1,1)$}$ is essentially the same as for the $SU(6)$ case, discussed in Sec.~\ref{sec:DomainWallsSU6}. Indeed, the physics in the two vacua is the same, a trivially gapped phase on one side and a $\mathbb{Z}_2$ theory on the other. It should then contain a 3d TQFT $\mathcal{A}^{2,1} = U(1)_2$, as we review briefly below.

In the present general case, the anomaly \eqref{eq:MixedAnomGenSUn} has $q=2$ and $a(N-1) = (2n+1) \times (4n+1)=1 \mod 2$. Hence, upon gauging the $\mathbb{Z}_2^{(1)}$ electric symmetry, the $\mathbb{Z}_{4n}$ axial symmetry defect $D$ needs to be decorated by the 3d TQFT
\begin{equation}
\mathcal{A}^{2,1} = U(1)_2 \, . 
\end{equation}
The coupling to the $\mathbb{Z}_2^{(1)}$ gauge field makes the domain wall no longer invertible. 
The story is exactly as for $SU(6)/\mathbb{Z}_2$, except that now $D^{4n}=1$ (always taking into account that $-1$ is a gauge transformation).

Again, note that the domain wall to go from the $j^{\textrm{th}}$ vacuum to the $(j+2)^{\textrm{th}}$ vacuum remains invertible since it does not have to host any TQFT. Hence, it simply consists of $D^2(M_3)$. 

A final comment concerning the generalization to symmetric matter. Nothing changes except the number of vacua, which is now $4n+4$.

\subsection{\texorpdfstring{$SU(4n)/ \mathbb{Z}_2$}{SU(4n)/Z2} for \texorpdfstring{$n \geq 2$}{n >=2}}
We finally consider the case of $\mathfrak{su}(4n)$, which ends up displaying the same physics as its first representative, $\mathfrak{su}(8)$.
(Indeed, the case of $\mathfrak{su}(4)$ is peculiar due to the reality of the anti-symmetric representation, as we have already discussed in detail.)

\subsubsection{Before gauging}
The $SU(4n) + \textrm{$\tiny\yng(1,1)$}$ theory with $n>1$ has a $\mathbb{Z}_{2(4n-2)}^{(0)}$ axial symmetry and a $\mathbb{Z}_2^{(1)}$ electric symmetry. One has $a=2n$ so that gcd$(a,q)=$ gcd$(2n,2) =2 \neq 1$. We end up with $q' = 1$, this means that there is no mixed 't Hooft anomaly between the axial symmetry and the electric symmetry.

Again, the fermion bilinear is assumed to condense in the IR and this spontaneously breaks the axial symmetry $\mathbb{Z}_{2(4n-2)}^{(0)} \rightarrow \mathbb{Z}_{2}^{(0)}$, yielding $4n-2$ isolated vacua. These vacua are related by the broken generators of $\mathbb{Z}_{4n-2}^{(0)} = \mathbb{Z}_{2(4n-2)}^{(0)} / \mathbb{Z}_{2}^{(0)}$, i.e. the first $4n-2$ rotations of $\mathbb{Z}_{2(4n-2)}^{(0)}$. All the vacua are physically equivalent.

The line operators of $SU(4n) + \textrm{$\tiny\yng(1,1)$}$ are purely electric. Due to the charged matter, the electric charges of the lines are defined mod 2. Thus the set of lines is 
\begin{equation}
SU(4n+2) +  \textrm{$\tiny\yng(1,1)$} \,: \{ (0,0), (1,0) \} \, . 
\end{equation}

\subsubsection{After gauging}

Now we gauge $\mathbb{Z}_2^{(1)}$ and we end up on the theory $SU(4n)/\mathbb{Z}_2 +  \textrm{$\tiny\yng(1,1)$}$. It has a $\mathbb{Z}_2^{(1)}$ magnetic one-form symmetry and a $\mathbb{Z}_{2(4n-2)}^{(0)}$ axial symmetry.

Again, two distinct theories share this gauge group. For pure gauge theories, $(SU(4n)/\mathbb{Z}_2)_0$ has lines which are  combinations of $(2,0)$ and $(0,2n)$, while $(SU(4n)/\mathbb{Z}_2)_1$ has combinations of $(2,0)$ and $(1,2n)$. With the fermions, we have:
\begin{equation}
\left( \frac{SU(4n)}{\mathbb{Z}_2} \right)_0 +  \textrm{$\tiny\yng(1,1)$} \, : \{ (0,0),(0,2n) \} \, , 
\end{equation}
\begin{equation}
\left( \frac{SU(4n)}{\mathbb{Z}_2} \right)_1 +  \textrm{$\tiny\yng(1,1)$} \,: \{ (0,0), (1,2n) \} \, .
\end{equation}
As expected from the fact that gcd$(a,q)=$gcd$(2n,2) \neq 1$, the two theories are not related by $\theta \rightarrow \theta + 2\pi$, and this shift is a symmetry within each of the two theories.

\subsubsection{Vacuum structure}

For the $SU(4n) +  \textrm{$\tiny\yng(1,1)$}$ theory, we assume as always that confinement  is due to the condensation of a dyonic particle in each vacuum, starting with the $(0,1)$ pure 't Hooft line in the $0^{\textrm{th}}$ vacuum. Acting with $\theta \rightarrow \theta + 2 \pi$ one then alternates between $(1,1)$ and $(0,1)$ by the Witten effect, so that we have 
\begin{align}
\textrm{Vacua } 0, 2,  4, ... ,  4n-4 & \rightarrow \langle T_{(0,1)} \rangle \neq 0 \, , \nonumber\\
\textrm{Vacua } 1,3,  5, ... , 4n-3 & \rightarrow \langle T_{(1,1)} \rangle \neq 0 \, .
\end{align}
The genuine $(1,0)$ line of the theory is never aligned with the condensing dyons in the vacua. Thus it exhibits an area law and the electric one-form symmetry is unbroken. The $4n-2$ vacua are trivially gapped, physically equivalent and related by an invertible symmetry.

Going to the $SU(4n)/\mathbb{Z}_2 +  \textrm{$\tiny\yng(1,1)$}$ theory, we see that since there is no mixed 't Hooft anomaly in the ungauged theory, the $\theta$-shifts to go from one of the $4n-2$ vacua to another are still invertible symmetries. The fate of the genuine line operators of each of the two theories is the following: 
\begin{itemize}
\item $\left( \frac{SU(4n)}{\mathbb{Z}_2} \right)_0 +  \textrm{$\tiny\yng(1,1)$}$ only has $(0,2n)$ that carries non-trivial charges. This line is aligned with both condensing lines $(0,1)$ and $(1,1)$ in the sense that $(0,2n) = (0,1)^{2n}$ and $(0,2n) = (1,1)^{2n}$, since the electric charge is mod-2 defined. Hence, in each of the $4n-2$ vacua, the IR theory is a $\mathbb{Z}_2$ gauge theory. 
\item $\left( \frac{SU(4n)}{\mathbb{Z}_2} \right)_1 +  \textrm{$\tiny\yng(1,1)$}$ only has $(1,2n)$ that carries non-trivial charges, but it is never aligned with the condensing line in any vacuum. The magnetic one-form symmetry remains unbroken and all $4n-2$ vacua are trivially gapped.
\end{itemize}
In both theories all the vacua are physically equivalent. This also confirms the fact that a $\theta$-shift is a symmetry within each theory, going from one vacuum to another. The same conclusions can be drawn in the case of symmetric matter, upon changing the number of vacua to $4n+2$.

\section{Non-invertible symmetries in \texorpdfstring{$\mathcal{N}=1$}{N=1} Super Yang-Mills \texorpdfstring{$USp(2N)/ \mathbb{Z}_2$}{USp(2N)/Z2}}
\label{sec:usp}
In this section we investigate theories with gauge algebra $\mathfrak{usp}(2N)$. Since the center of the simply connected group $USp(2N)$ is $\mathbb{Z}_2$, we do not have to look far for a representation that preserves it, namely the adjoint is perfectly suited. Since the adjoint is real, a Weyl fermion is enough to carry it, and we are hence left with a $\mathcal{N}=1$ supersymmetric gauge theory. We have then a better control over the number of vacua, via for instance the Witten index. Other than that, the following discussion is independent on supersymmetry.

\subsection{Before gauging}
We work with $USp(2N) = Sp(N)$, where $USp(2N)$ is a compact, simply connected group of real dimension $N(2N+1)$ and has $\mathbb{Z}_2$ as a center subgroup. The Lie algebra $\mathfrak{usp}(2N)$ has rank $N$ and all its irreducible representations are real or pseudo-real. They have $N$-ality $c = 0$ or $c=1$, since it is defined mod $2$.\footnote{Hence it would perhaps be more appropriate, but confusing, to call it 2-ality. For an exhaustive list of properties of the representations of $\mathfrak{usp}(2N)$, see \cite{Yamatsu:2015npn}.} 
Some characteristics of the simplest irreducible representations of $USp(2N)$ can be found in Tab.~\ref{table=USp}. Note that the adjoint representation is nothing but the two-index symmetric representation. 

\begin{table}
\centering
\begin{tabular}{||c c c c c||} 
 \hline
 Irrep & Dynkin labels & Dimension & Dynkin index & $N$-ality \\ [0.5ex] 
 \hline
 \textrm{$\tiny\yng(1)$} & (1,0,...,0) & $2N$ & 1 & 1 \\ [0.5ex]
 \hline
 \textrm{$\tiny\yng(1,1)$} & (0,1,0,...,0) & $N(2N-1)-1$ & $2N-2$ & 0 \\ [0.5ex]
 \hline
 \textrm{$\tiny\yng(2)$} & (2,0,...,0) & $N(2N+1)$ & $2N+2$ & 0 \\ [0.5ex] 
 \hline
\end{tabular}
\caption{Some irreducible representations of $\mathfrak{usp}(2N)$}
\label{table=USp}
\end{table}

Let us take one real Weyl fermion $\lambda$ in the adjoint representation, in order to realize  $\mathcal{N}=1$ supersymmetry. 
The beta function at one-loop is given by 
\begin{align}
\beta (g) &= 
- 3 (N+1)  \frac{g^3}{16 \pi^2} \, ,
\end{align}
ensuring that the theory is also asymptotically free for any value of $N$.
The $\mathcal{N}=1$ SYM $USp(2N)$ theory has a $U(1)_R^{(0)}$ R/axial-symmetry which is again ABJ-anomalous, with a discrete $\mathbb{Z}_{2(N+1)}^{(0)}$ that survives quantum mechanically. In the IR, it is spontaneoulsy broken to $\mathbb{Z}_2$ by a gaugino condensate $\langle \tr ( \lambda \lambda ) \rangle \neq 0$, yielding $N+1$ distinct vacua that are related by the broken generators of $\mathbb{Z}_{N+1} = \mathbb{Z}_{2(N+1)}/\mathbb{Z}_2$, or equivalently by shifts of $\theta$. They are distinguished by the value of the gaugino condensate
\begin{equation}
\langle \tr ( \lambda \lambda ) \rangle  = \Lambda^3  e^{\frac{2\pi i  k}{N+1}} \, , ~ ~ \textrm{with } k=0,1, ..., N  \, . 
\end{equation}
It also has a $\mathbb{Z}_2^{(1)}$ electric symmetry related to the center of the gauge group. In addition to the trivial line of charge $(0,0)$, the theory only contains the purely electric line $(1,0)$.

There is a mixed 't Hooft anomaly between the axial/R-symmetry of the fermion and the electric one-form symmetry. Following \cite{Cordova:2019uob}, the anomaly between the transformation $\theta \rightarrow \theta + 2\pi n$ and the $\mathbb{Z}_2^{(1)}$ is 
\begin{equation}\label{eq:UspAnomaly}
A = 2\pi n \,  \frac{N}{4} \int \mathcal{P}[\hat{B}_e]
\end{equation}
where $\hat{B}_e$ is the 2-cocycle associated to the 2-form background field of $\mathbb{Z}_2^{(1)}$, namely $\hat{B}_e \in H^2(X, \mathbb{Z}_2)$. On spin manifolds, the integral of the Pontryagin square is an even number, so that the anomaly ends up being at most $\mathbb{Z}_2$-valued. It is trivial for $N$ even, implying that there is no anomaly. However, one half of the axial rotations are anomalous for odd $N$.

\subsection{After gauging}\label{sec:N1USp/Z2}

In the $USp(2N)/ \mathbb{Z}_2$ theory, instead of the $(1,0)$ line, one can either have the purely magnetic line $(0,1)$ or the dyonic line $(1,1)$. These two possibilities are related to two distinct theories, respectively $(USp(2N)/ \mathbb{Z}_2)_+$ and $(USp(2N)/ \mathbb{Z}_2)_-$, that share the same local physics but are different when it comes to global features.

The consequences of a $2\pi$-shift of $\theta$ in these theories have been studied in \cite{Aharony:2013hda}. Not surprisingly, the $\theta$-shift maps one theory to the other when the anomaly is non-trivial, while it is a true symmetry when there is no anomaly. We can summarize the results as follows:
\begin{itemize}
\item for $N$ even, the line with non-zero magnetic charge remains as it is after the $2\pi$-shift of $\theta$. It implies that $(0,1)$ remains $(0,1)$ and $(1,1)$ remains $(1,1)$. One has 
\begin{equation}
(USp(2N)/ \mathbb{Z}_2)^\theta_\pm = (USp(2N)/\mathbb{Z}_2)^{\theta + 2\pi}_\pm ~\textrm{for even } N \, , 
\end{equation}
\item for $N$ odd, the line with non-zero magnetic charge gets an electric charge under the $2\pi$-shift of $\theta$. It implies that $(0,1)$ becomes $(1,1)$ and $(1,1)$ becomes $(0,1)$. It yields
\begin{equation}\label{eq:cecicela}
(USp(2N)/ \mathbb{Z}_2)^\theta_\pm = (USp(2N)/\mathbb{Z}_2)^{\theta + 2\pi}_\mp ~\textrm{for odd } N \, . 
\end{equation}
\end{itemize}

\subsection{Vacua}

In the $USp(2N)$ theory, there are $N+1$ distinct vacua that are physically equivalent and related by the genuine symmetries $\theta \rightarrow \theta + 2 \pi n$. One expects that confinement in all the $N+1$ vacua is due to the condensation of a magnetically charged particle, in such a way that the line $(1,0)$ is never aligned with this condensing particle in any vacuum and it exhibits an area law. Let us assume that $(0,1)$ condenses in the $0^{\textrm{th}}$ vacuum. Now one needs to figure out what happens for the other vacua. Notice that the previous discussion of Sec.~\ref{sec:N1USp/Z2} about the effect of $\theta$-shifts on the dyons holds irrespectively of the global form of the gauge group, and hence also for the $USp(2N)$ theory. Again, we shall split the story into two parts. 

\subsubsection{Even $N$}
In this case, shifting $\theta$ does not affect the lines. Accordingly, we take to be the line $(0,1)$ that condenses in all vacua of the $\mathcal{N}=1$ $USp(2N)$ theory.
This agrees with the fact that there is an odd number of vacua, meaning that we could not come back to $(0,1)$ in the $0^{\textrm{th}}$ vacuum if the Witten effect would have been non-trivial. Eventually, there is confinement in all vacua and the low-energy theory is trivially gapped.

In the gauged theory $ USp(2N)/ \mathbb{Z}_2$, there are still $N+1$ vacua and the same magnetic particle $(0,1)$ condenses in each one of them.
Since there is no mixed 't Hooft anomaly (\ref{eq:UspAnomaly}) in the ungauged theory, $\theta \rightarrow \theta + 2 \pi n$ remain genuine symmetries relating the vacua, that are still physically equivalent. In $(USp(2N)/ \mathbb{Z}_2)_+$, the non trivial line operator of the theory $(0,1)$ is obviously aligned with the condensing particle, hence it exhibits a perimeter law and the low-energy theory is a $\mathbb{Z}_2$ gauge theory in all vacua. In $(USp(2N)/ \mathbb{Z}_2)_-$, there is no alignment and the non trivial line operators keep their area law, so that in all vacua the theory is trivially gapped.

\subsubsection{Odd $N$}
Here, the Witten effect takes place as usual and this implies that the dyonic particle $(k \mod 2 ,1)$ condenses in the $k^\textrm{th}$ vacuum of the $USp(2N)$ theory. As expected, it comes back to $(0,1)$ after going through the total (even) number of vacua. The purely electric line of the theory still exhibits an area law in all vacua. The theory is trivially gapped in every vacuum, though as usual the vacua are two by two equivalent at the level of the SPT phases.

In the gauged theory $ USp(2N)/ \mathbb{Z}_2$, the same dyons must condense in their respective vacuum. However in this case there is a non-trivial mixed 't Hooft anomaly (\ref{eq:UspAnomaly}) in the ungauged theory, implying 
\begin{equation}
\theta \rightarrow \theta + 2\pi n \, \,   \left\{
    \begin{array}{ll}
        \textrm{is a symmetry} & \textrm{for  }  n=0,2, ... ,N-2 \\
        \textrm{is a non-invertible symmetry} & \textrm{for  }  n=1,3, ... , N \, . 
    \end{array}
\right.
\end{equation}
Thus, one expect two adjacent vacua to be physically inequivalent, since they are not related by a genuine invertible symmetry. Indeed, in $(USp(2N)/ \mathbb{Z}_2)_+$, the non trivial line operators of the class $(0,1)$ are aligned with the condensing particle $(0,1)$ in half of the vacua, namely the vacua that are labelled by an even number. There is no alignement in the other half. At the end, going to low energies yields a $\mathbb{Z}_2$ gauge theory in the even vacua and a trivially gapped theory (with an SPT$_1$) in the odd vacua. It is the converse for $(USp(2N)/ \mathbb{Z}_2)_-$, where one has a trivially gapped theory in the even vacua and a $\mathbb{Z}_2$ gauge theory in the odd vacua. All in all, this agrees with (\ref{eq:cecicela}), in the sense that a $2\pi$-shift of $\theta$ both changes vacuum and the theory. The vacua of $(USp(2N)/ \mathbb{Z}_2)_+$ and $(USp(2N)/ \mathbb{Z}_2)_-$ are the same, provided that one relabels them.

The difference in the anomaly between two adjacent vacua is $2\pi \frac{N}{4} \int \mathcal{P} [\hat{B}]$, where $N$ is an odd number. Hence, one should decorate again the basic domain wall with the minimal 3d TQFT $\mathcal{A}^{2,1}$. The discussion is the same as in the cases analyzed before, in the sense that the non-invertible domain wall interpolates between vacua whose characteristic is to be trivially gapped on one side, and a $\mathbb{Z}_2$ gauge theory on the other. The non-invertibility of the domain wall is related to its coupling to the 2-form gauge field implementing the $\mathbb{Z}_2$ quotient of the gauge group. Other details of the gauge theory do not matter. For instance, the fusion rules of the defects are exactly the same as for $SU(4n+2)/\mathbb{Z}_2 +  \textrm{$\tiny\yng(1,1)$}$, which we exemplified for $n=1$ in \S \ref{sec:DomainWallsSU6}. 

Note that the same TQFT can be obtained by different arguments, pertinent to ${\cal N}=1$ SYM with generic gauge group. In \cite{Delmastro:2020dkz}, it is found that for gauge group $USp(2N)$, the TQFT on the domain wall between adjacent vacua should be $USp(2)_N$ CS theory. Now the latter theory, seen as $SU(2)_N$, is level-rank dual to $U(N)_{-2}$. For odd $N$, the $\mathbb{Z}_2$ one-form symmetry of this CS theory has a 't Hooft anomaly \cite{Gaiotto:2014kfa}, which is actually captured by its $\mathcal{A}^{2,1}$ TQFT sector (recall that $\mathcal{A}^{2,1} = U(1)_{2}\simeq U(1)_{-2}$). Hence in the $USp(2N)/ \mathbb{Z}_2$ theory with odd $N$, the minimal TQFT coupling to the 2-form gauge field coincides with what we found.

\section{Comments on theories with \texorpdfstring{$\mathfrak{so}(N)$}{so(N)} gauge algebra}
\label{sec:spin}

We will not go into much detail for this case. It decomposes into some subcases. First of all for $N$ odd, the simply connected group $Spin(2n+1)$ has a $\mathbb{Z}_2$ center. However it is shown in \cite{Cordova:2019uob} that the anomaly under $2\pi$ shifts of $\theta$ is always trivial on spin manifolds. Hence we do not expect any non-invertible symmetries upon going to the two $Spin(2n+1)/\mathbb{Z}_2=SO(2n+1)$ theories. For even $N$, one should make the distinction between $Spin(4n+2)$ which has a $\mathbb{Z}_4$ center and $Spin(4n)$ which has a $\mathbb{Z}_2\times \mathbb{Z}_2$ center. According to \cite{Cordova:2019uob} the anomaly is now non-trivial also on spin manifolds. However, if there is dynamical charged matter in the vector representation, these electric one-form symmetries are broken to $\mathbb{Z}_2^{(1)}$ and it turns out that the anomaly becomes trivial on spin manifolds. Consequently there is no non-invertible symmetry defect in any of the $Spin(N)/\mathbb{Z}_2=SO(N)$ theories.\footnote{The $SO$ theory is obtained by gauging the $\mathbb{Z}_2$ subgroup of the center $\mathbb{Z}_4$ in $Spin(4n+2)$ and the diagonal $\mathbb{Z}_2$ subgroup of the center $\mathbb{Z}_2\times \mathbb{Z}_2$ in $Spin(4n)$.}

Let us now as a final case go slightly beyond the main object of this paper, and gauge more than a $\mathbb{Z}_2$ subgroup of the one-form electric symmetry. For instance, for $N$ even, one could gauge the full center symmetry if the matter is in the adjoint representation. Let us stick to studying $\mathcal{N}=1$ SYM $Spin(4n+2)/\mathbb{Z}_4 $ theory for the sake of simplicity.

The $\mathcal{N}=1$ SYM $Spin(4n+2)$ theory has a $\mathbb{Z}_{8n}^{(0)}$ axial/R-symmetry which is spontaneously broken to $\mathbb{Z}_{2}^{(0)}$ due to the bilinear gaugino condensate, yielding $4n$ distinct vacua. The mixed axial-electric 't Hooft anomaly \cite{Cordova:2019uob} is given by 
    \begin{equation}\label{eq:AnomalySpin4n+2}
        A = 2\pi m \frac{4n+2}{16} \int \mathcal{P}[\hat{B}_e]
    \end{equation}
under $\theta \rightarrow \theta + 2 \pi m$. It is $\mathbb{Z}_4$-valued and the only non-anomalous axial rotation angles are characterized by $m$ being an integer multiple of $4$.
The only line operators in the theory are the purely electric ones generated by $(1,0)$. In the IR, the $4n$ vacua are trivially gapped even if they are divided into 4 classes that differ by their SPT phase, due to the mixed anomaly being valued in $\mathbb{Z}_4$. One assumes that the theory confines due to condensation of dyonic lines. Starting with the condensation of $(0,1)$ in the $0^\textrm{th}$ vacuum, the other condensing lines are obtained via the Witten effect. In particular, the $k^\textrm{th}$ vacuum is characterized by the condensation of $(k \mod 4,1)$.

The $\mathcal{N}=1$ SYM $Spin(4n+2) / \mathbb{Z}_4$ is obtained by gauging the electric one-form symmetry of the $Spin(4n+2)$ theory. Doing so, one ends up with axial transformations being such that
    \begin{equation}
    \theta \rightarrow \theta + 2\pi m \, \,   \left\{
\begin{array}{ll}
        \textrm{is a symmetry} & \textrm{for  }  m=0 \mod 4 \\
        \textrm{is a non-invertible symmetry} & \textrm{for  }  m=1,2,3 \mod 4 \, . 
    \end{array}
    \right.
    \end{equation}
There are 4 distinct theories that share the $Spin(4n+2) / \mathbb{Z}_4$ gauge group but that differ in their line operator content \cite{Aharony:2013hda}. They specifically are labelled such that  
    \begin{equation}
        \left( \frac{Spin(4n+2)}{\mathbb{Z}_4} \right)_p \, : \, \{ (0,0) , (p,1) \}  \, . 
    \end{equation}
Note that the Witten effect depends on $n$ since  $(z_e, z_m) \rightarrow (z_e \pm z_m , z_m)$ under $\theta \rightarrow \theta + 2 \pi$, where the $+$ is associated with $n$ odd and the $-$ with $n$ even. Hence the $2\pi$-shift of $\theta$ goes from one theory with $p$ to the next/previous one with $p \pm 1$. There are still $4n$ vacua, that are associated to the same condensing lines as in the ungauged theory. However, genuine lines of the theory can now be aligned with these condensing dyons and thus have to share their perimeter law. This actually results in 3 classes of physically distinct vacua. Let us take $(Spin(4n+2) / \mathbb{Z}_4)_0$.
\begin{itemize}
    \item  In the $0^\textrm{th}$ vacuum, its genuine line $(0,1)$ is obviously aligned with the condensing line $(0,1) \sim (0,1)$, hence has a perimeter law yielding a $\mathbb{Z}_4$ gauge theory in this vacuum. 
        \item In the $1^\textrm{st}$ vacuum, the non-trivial line has an area law and the theory is trivially gapped.
    \item In the $2^\textrm{nd}$ vacuum, one has $(0,1)^2 \sim (2,1)^2$, yielding a $\mathbb{Z}_2$ gauge theory.
    \item In the $3^\textrm{rd}$ vacuum, the theory is again trivially gapped. 
\end{itemize}
This pattern is repeated $n$ times among the $4n$ vacua.
 The story is the same in the 3 other $Spin(4n+2) / \mathbb{Z}_4$ theories provided that one relabels the vacua, according to the consequences of $\theta \rightarrow \theta + 2 \pi$. Note that the two classes of vacua that are trivially gapped are not related by an invertible symmetry due to the shift in the anomaly between the two sides of the corresponding domain wall. Also, they differ by the SPTs inherited from the $Spin(4n+2)$ theory. More specifically, in order to account for the shifts of (\ref{eq:AnomalySpin4n+2}) with $m=1,2,3$, one finds that
 \begin{enumerate}
     \item The basic domain wall separating two adjacent vacua ($m=1$) must host as minimal abelian 3d TQFT an $\mathcal{A}^{4,3} = SU(4)_1 \simeq U(1)_{-4}$ for odd $n$ and an $\mathcal{A}^{4,1} = U(1)_4 $ for even $n$.
     \item The domain wall with $m=2$ has to be decorated with an $\mathcal{A}^{2,1}$ topological theory.
     \item The domain wall with $m=3$ has to host an $\mathcal{A}^{4,1}  $ for odd $n$ and an $\mathcal{A}^{4,3}  $ for even $n$.
 \end{enumerate}
The fact that the domain walls for $m=1,3$ are reversed for even $n$ compared to odd $n$ agrees with the Witten effect going the other way around in both cases.

Finally, we note that the classes of vacua and the domain walls among them are essentially independent on $n$, i.e.~the rank, and are thus physically equivalent to those obtained in the $n=1$ case, which is nothing else than $\mathcal{N}=1$ SYM with gauge group $PSU(4) = SU(4) / \mathbb{Z}_4$ \cite{Choi:2022zal}. Accordingly, we expect the non-invertible fusion rules of the defects to be rank-independent. Note however that these display some subtleties, most notably in the fusion of two basic defects, as discussed in \cite{Copetti:2023mcq}.

\subsection*{Acknowledgements}
We thank Jeremias Aguilera Damia, Pierluigi Niro and Luigi Tizzano for many helpful discussions. R.A.~and R.V.~are respectively a Research Director and a Research Fellow of the F.R.S.-FNRS (Belgium). This research  is supported by IISN-Belgium (convention 4.4503.15) and through an ARC advanced project.


\bibliographystyle{JHEP}
\bibliography{library2}

\end{document}